\documentclass[acmsmall]{acmart}

\def\BibTeX{{\rm B\kern-.05em{\sc i\kern-.025em b}\kern-.08emT\kern-.1667em\lower.7ex\hbox{E}\kern-.125emX}}

\usepackage[utf8]{inputenc}
\usepackage{url}
\usepackage{xcolor}
\usepackage{comment}
\usepackage{multirow}
\usepackage{tikz} 
\usepackage{multirow}
\usepackage{makecell}
\usepackage{float}

\usepackage{booktabs}

\newcommand*\circled[1]{\tikz[baseline=(char.base)]{
            \node[shape=circle,draw,inner sep=1pt] (char) {\scriptsize #1};}}
\newcommand*{\M}{\mathcal{M}}
\newcommand*{\ReKeyGen}{\mathop{\mathsf{ReKeyGen}}}
\newcommand*{\ReEnc}{\mathop{\mathsf{ReEnc}}}
\newcommand*{\Verify}{\mathop{\mathsf{Verify}}}
\providecommand{\keywords}[1]{\textbf{\textit{Keywords---}} #1}

\newcounter{mylabelcounter}

\makeatletter
\newcommand{\labelText}[2]{
#1\refstepcounter{mylabelcounter}
\immediate\write\@auxout{
  \string\newlabel{#2}{{1}{\thepage}{{\unexpanded{#1}}}{mylabelcounter.\number\value{mylabelcounter}}{}}
}
}
\makeatother

\usepackage{xspace}
\usepackage{comment}
\newcommand{\heading}[1]{{\vspace{6pt}\noindent\sc{#1.}}} 
\mathchardef\mhyphencar ="2D

\newcommand{\ID}{\id}

\newcommand{\secpar}{\lambda}
\newcommand{\secparam}{1^\secpar}

\newcommand{\ptxt}{\mathit{M}}
\newcommand{\ctxt}{\mathsf{CT}}

\DeclareMathOperator{\sample}{{\leftarrow\!\!\mbox{\tiny${\$}$\normalsize}}\,}

\newcommand{\sk}{\mathsf{sk}}

\newcommand{\Enc}{\mathsf{Enc}}
\newcommand{\Dec}{\mathsf{Dec}}

\newcommand{\pk}{\mathsf{pk}}

\newcommand{\Sign}{\mathsf{Sign}}

\newcommand{\id}{\mathsf{id}}

\newcommand{\Setup}{\mathsf{Setup}}

\newcommand{\mpk}{\mathsf{mpk}}
\newcommand{\msk}{\mathsf{msk}}

\newcommand{\setup}{\mathsf{Setup}}
\newcommand{\keygen}{\mathsf{KeyGen}}

\newcommand{\KeyGen}{\mathsf{KeyGen}}

\begin{document}
\title{Security and Privacy in Big Data Sharing: State-of-the-Art and Research Directions}

\author{Houda Ferradi}
\affiliation{
\institution{The Hong Kong Polytechnic University}
\streetaddress{Hung Hom}
\city{Hong Kong}
\country{Hong Kong}}

\author{Jiannong Cao}
\affiliation{
\institution{The Hong Kong Polytechnic University}
\streetaddress{Hung Hom}
\city{Hong Kong}
\country{Hong Kong}}

\author{Shan Jiang}
\affiliation{
\institution{The Hong Kong Polytechnic University}
\streetaddress{Hung Hom}
\city{Hong Kong}
\country{Hong Kong}}
 
\author{Yinfeng Cao}
\affiliation{
\institution{The Hong Kong Polytechnic University}
\streetaddress{Hung Hom}
\city{Hong Kong}
\country{Hong Kong}}

\author{Divya Saxena}
\affiliation{
\institution{The Hong Kong Polytechnic University}
\streetaddress{Hung Hom}
\city{Hong Kong}
\country{Hong Kong}}

\begin{abstract}

Big Data Sharing (BDS) refers to the act of the data owners to share data so that users can find, access and use data according to the agreement. In recent years, BDS has been an emerging topic due to its wide applications, such as big data trading and cross-domain data analytics. However, as the multiple parties are involved in a BDS platform, the issue of security and privacy violation arises. There have been a number of solutions for enhancing security and preserving privacy at different big data operations (e.g., data operation, data searching, data sharing and data outsourcing). To the best of our knowledge, there is no existing survey that has particularly focused on the broad and systematic developments of these security and privacy solutions. In this study, we conduct a comprehensive survey of the state-of-the-art solutions introduced to tackle security and privacy issues in BDS. For a better understanding, we first introduce a general model for BDS and identify the security and privacy requirements. We discuss and classify the state-of-the-art security and privacy solutions for BDS according to the identified requirements. Finally, based on the insights gained, we present and discuss new promising research directions.

\end{abstract}
\maketitle     

\keywords{Big data sharing, Security, Privacy, Data Ownership, Server, Blockchain}

\section{Introduction}
\label{Introduction}

The term \emph{big data} as the name suggest refers to information assets characterized by high volume, fast access speed, and a large ontological variety. Dealing with big data requires specific technologies and analytical methods for its transformation into value. The term \emph{big data sharing} (BDS) refers to the act of the data sharer to share big data so that the data sharee can \emph{find, access}, and \emph{use} in the agreed ways. BDS not only improves the speed of getting data insights, but can also help strengthen cross-domain data analytics and big data trading. Over the last few years, there is a huge demand for big data sharing in various industries, which has led to an explosive growth of information. Over 2.5 quintillion bytes of data are created every single day, and the amount of data is only going to grow from there. By 2020, it is estimated that 1.7MB of data will be created every second for every person on earth. Due to constraints related to the limitations of data storage resources, large storage dedicated centralized servers (e.g., cloud) are usually regarded as the best approach for BDS: on the one hand the centralized server provides a solution which is both scalable and accommodating for BDS and business analytics,  while on the other hand BDS provides data analytics for actionable insight and making predictions. However, centralized server comes with a price: it constitutes an added level of security and privacy threats since its essential services are often outsourced to an untrusted third party, which makes harder to maintain the basic data security and privacy requirements, such as confidentiality, integrity and privacy of the shared data. Thus, enforcing security and privacy in BDS as a whole is an important concern. Otherwise, data integrity and confidentiality can always be compromised easily. 

\subsection{General Model of BDS and its Security and Privacy Concerns}
\label{SystemModel}

In this section we first discuss the general model of BDS. Then, we describe the general operations needed based on that general model. According to that, we broadly categorize the security and privacy notions needed for BDS.

\textbf{General Model of BDS.} Before discussing security and privacy concerns,  it is necessary to define a general model of BDS and its input/output. The general model that we consider is based on the centralized server approach. We propose a general model of BDS system that allows data sharer/sharee to \emph{create, store, access, download, search} and \emph{manipulate databases}, that takes full account of data access control, user accessibility and the form of shared data. 
    
The remote centralized service provider (e.g., Cloud) which stores and manages
the data generated by data sharer is considered as an untrusted party by the two other parties. The sharing activity could be either operating on data, e.g., searching or computation or downloading data. The shared data could be either raw or encrypted. Table ~\ref{notations} shows the notations.

\begin{figure}[!htbp]
\includegraphics[width=.75\linewidth]{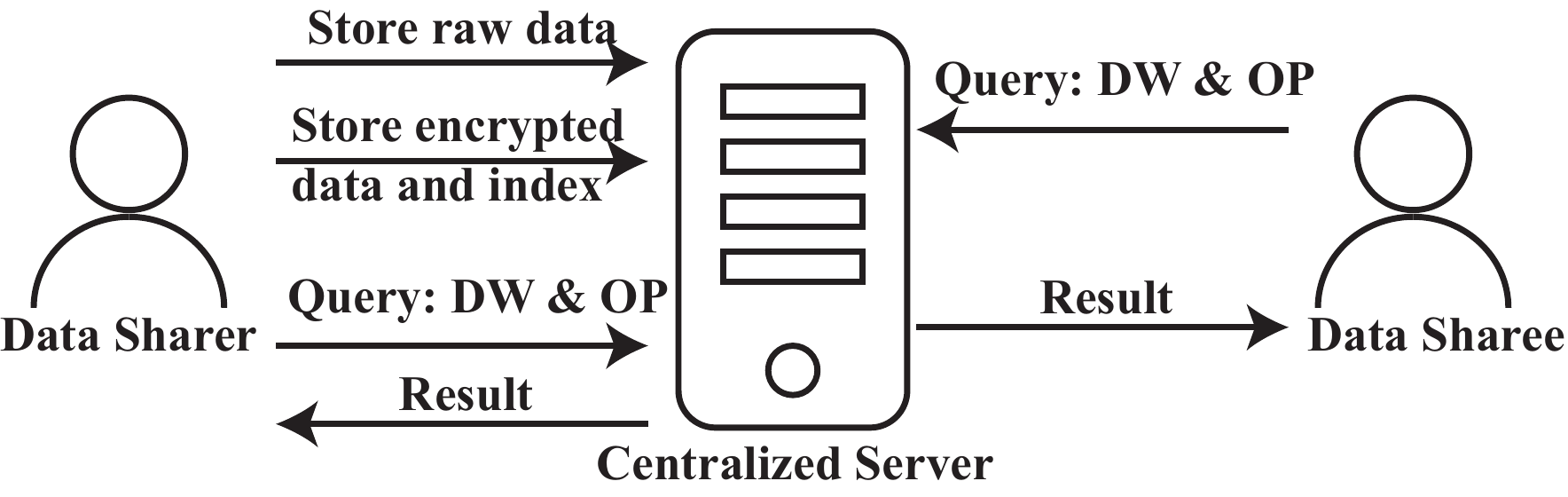}
\caption{General Model for BDS}
\label{fig:centralized model}
\end{figure}

\begin{table*}[tbp]
\centering
\caption{Notations} \label{notations}
\scalebox{0.65}{\begin{tabular}{ p{.15\linewidth} p{.58\linewidth}  }
\toprule
Notations   & Descriptions 
\\ \toprule
$\mathsf{DW}$ &  Downloaded data \\ \midrule
$\mathsf{OP}$ & Data operations which includes: data searching, data outsourcing, data computation \\ \midrule
$\mathsf{ENC}$ & Data is encrypted during the storage and sharing process \\ \midrule 
$\mathsf{RAW}$ & Data is raw during the storage and sharing process \\ \midrule
$\mathsf{INT}$ & Data Sharer/Internal Use \\ \midrule
$\mathsf{EXT}$ &  Data Sharee/External User \\ 
\bottomrule
\end{tabular}}
\end{table*}

Our general model for BDS consists of the two following entities (as shown in Fig. \ref{fig:centralized model}):

\begin{description}
     \item[Data Sharer/Internal User.]  A data sharer is the data owner (or internal user) who shares his own data with a larger server storage. In such system, the data sharer can either use and operate on his own shared data or gives its access to data sharee.
     \item[Data Sharee/External User.] A data sharee (or external user) access/uses other's stored data. In such system, the data sharer gives the access to that data to the data sharee, either by downloading data from the server or by directly operating on data.
\end{description}

After defining what is BDS made of, we explain its general procedures.

BDS operations can be divided into a few distinct groups, which have their own characteristics. Herein, we define and discuss the general operations needed for BDS, i.e., data downloading, data storing, data computation, data searching, and data outsourcing:
 
\textsc{Data sharing.} Data sharer might query on a BDS platform with some constraints, to learn hidden patterns, correlations, compute a function, and other insights. Generally speaking, there are three steps for data querying,
i.e., data computation, data downloading, data searching, and data outsourcing, as follows:

\begin{itemize}
    \item \textsc{Data downloading.}  Data downloading is the process through which data sharer or data sharee retrieve the result from data querying. 
    \item \textsc{Data operation.}
    \begin{enumerate}
        \item \textsc{Data computation.} Data computation  is the process through which data sharer or data sharee jointly compute a function with BDS over their inputs while keeping those inputs private.
        \item \textsc{Data searching.}  Data users might query a BDS platform with some constraints, to learn hidden patterns, correlations and other insights. Data searching needs to be able to search through unstructured and structured data which requires management of huge amounts of data as quickly as possible.
        \item \textsc{Data outsourcing.} Data users might delegate a portion of data to be outsourced to external providers who offer data management functionalities.
   \end{enumerate}
\end{itemize}
  
Although there are numerous benefits for BDS, it is non-trivial to design a solution because of the
requirements. So, security and privacy are necessary for BDS, otherwise its values will be disappeared, i.e.,
if a BDS is not secure or cannot protect
the security and privacy, then the users can hardly trust such a technology and
even will not use it. Below, we categorize and explain the
basic concerns, i.e., \emph{data security}, \emph{data privacy}, and \emph{user privacy} as follows:

\textbf{Security and Privacy Concerns in BDS.} Considering these data sharing operations, we categorize data security and privacy notions needed in BDS, i.e., Data Security, Data Privacy and User Privacy.
Data security refers to \emph{how} data is protected from an attacker, namely: Prevent malicious access, usage, modification or unavailability of the big data from anyone other than the sharing parties. 
Data Privacy is about protection of individual's information from being disclosed to others as data may contain individual's sensitive information, such as Personally Identifiable Information (PII), personal healthcare information, and financial information, which should be protected whenever the data is collected, stored and shared (e.g., by applying governing regulation or law like General Data Protection Regulation (GDPR)). There are two parties in BDS, which are the data sharer and sharee. User privacy is about protecting the identity of data sharer from exposure by other parties and even each other. It requires that the two parties involved in BDS focus on the data itself without knowing each other.

\subsection{Motivation and Contributions} For the past few years, the topic of big data security and privacy have been explored in many surveys. Most of these survey papers \cite{Thangaraj17,Siddique18,Terzi15,Chandra17,Fang17} give a short overview on security and privacy techniques in BDS.  This work aims to contribute a comprehensive survey of security and privacy in BDS in terms of formal definitions, security and privacy requirements, security and privacy techniques used to fulfill requirements, classification of techniques and future challenges.

Compared to other surveys that can be found in the literature, our contributions are as follows:

\begin{itemize}
    \item \textbf{New taxonomy.}  After providing an in-depth understanding and up-to-date discussion related to the BDS and its operations. We identify security and privacy requirements within BDS and present a novel taxonomy to structure solutions by fulfilled requirements. 
    \item \textbf{Comprehensive survey.} In accordance with the taxonomy, we discuss the benefits and limitations of the state of-the-art solutions that fulfill the identified security and privacy requirements.  
    \item \textbf{Future directions.} Finally, based on our survey, we provide  the list of lessons learned, open issues, and directions for future work.
\end{itemize}

\subsection{Security \& Privacy Concerns in Big Data Sharing Applications} 

BDS has many applications fields, such as healthcare \cite{poldrack2014making}, supply chain management \cite{richey2016global}\cite{wu2019data}, and open government \cite{welch2016determinants}. In this section, we introduce three attracting applications that caught the attention of both the industry and academia in recent years.

\textbf{Privacy and Pandemic.}
Global leaders are increasingly relying on information about individuals and communities  to control the spread of COVID-19 and respond to its economic, political, social, and health impacts. Time is of the essence, and leaders must quickly decide essential questions about what personal information they will collect or disclose, to whom, and under what conditions. It is important that privacy concerns do not become an obstacle to effective health and safety measures, but also that we do not open a door to privacy violation or limitless surveillance.

\textbf{Federated Learning (FL).}
FL is a subset within the field of AI, enables  multiple decentralized edge devices or servers  holding local data samples to collaboratively learn a shared prediction model while keeping all the training data private.
In recent years, FL has received extensive attention from both academia and industry because it can solve privacy problems in machine learning.  However, there are many challenges in FL, and although there are solutions to these challenges, most existing solutions need a trusted, centralized authority that is difficult to find.

\textbf{Medical Research and Healthcare.} In recent years, more and more health data are being generated. All these big data put together can be used to predict the onset of diseases so that preventive steps can be taken. However, the health data contains personal health information (PHI), due to the risk of violating the privacy there will therefore be legal concerns in accessing the data \cite{Jain16}. Health data can be anonymized using masking and de-identification techniques, and be disclosed to the researchers based on a legal data sharing agreement \cite{Khaled13}. 

\subsection{Organisation} This paper is structured as follows: In Section \ref{Introduction}, we start by defining BDS, this allows us to discuss the differences between the security and privacy notions in BDS. Next, we provide a comprehensive topical overview of BDS by introducing its general model and general procedures using centralized architecture. Based on this,  we describe the different assumptions and scope for that model. In Section \ref{sec:security}, we start by describing the basic security requirements as well as additional ones that are needed in BDS and then describing their corresponding techniques. In Section \ref{sec:privacy}, we describe the privacy requirements in terms of data and user privacy and their corresponding techniques. In Section \ref{Taxonomy security and privacy techniques}, we review, summarize and compare the security \& privacy techniques to fulfill the needed security \& privacy requirements. In Section \ref{Future Directions}, we discuss the challenge issues as well as new future research directions for BDS. Finally in Section \ref{conclusion}, we conclude this article.

\section{Security in BDS}
\label{sec:security}
In this section, we first start defining the required security properties  as well as setting the security assumptions. Based on this, we overview the existing cryptographic techniques. It allows us to describe how these techniques can be incorporated in the BDS system. Finally, we provide a classification that compare the various cryptographic techniques.

\subsection{Security Requirements in BDS}
\label{Security requirements}

In this section, we first start recalling the four most fundamental security requirements coming from information system, also known as
the CIA triad, which are defined as follows:

\heading{\textbf{Data Confidentiality during Outsourcing.}}
Confidentiality is the cornerstone of BDS security which refers to the protection of data during the sharing process against the unauthorized access.  Otherwise, its value could be disappeared.

\heading{\textbf{Data Integrity.}} We distinguish two types of integrity in data sharing context: \emph{Usage Integrity} (or Data Integrity) ensuring that any unauthorized modification of sensitive data in the use should be detectable, otherwise its veracity cannot be consistent. While, \emph{Data Source Authenticity} means that the data should be consistent over the whole BDS process. The distinction done between data integrity and authentication is frequently blurred because integrity can also provide authentication. In essence, an integrity primitive would take as a parameter a message $m$ and prove that the sender actually mixed his secret with $m$ to attest $m$'s origin. An authentication primitive does not involve any message (no "payload") and is only meant to check that the authenticated party actually knows a given secret. It follows that to achieve authentication the secret owner can just be challenged to attest the integrity of random challenge $m$, chosen by the verifier. In practice, this is indeed the way in which numerous commercial products implement authentication using integrity primitives.

\heading{\textbf{Non-repudiation}}
While integrity ensures a data has not been tampered with, non-repudiation provides evidence that an individual or entity from denying having performed a particular action. In other words, non-repudiation provides proof of the origin of data and the integrity of the data. 

\heading{\textbf{Availability}} Data availability ensures that data must be available for use whenever authorized users want it. However, the introduction of cloud computing has limited issues of data availability for Big Data due to high has narrowed down issues of cloud. Denial of service (DoS) attack, DDoS attack, and SYN flood attack are the most common attacks to threat data availability. 

Besides the basic security requirements of BDS, we specify the additional security requirements that we have identified for BDS context, which are defined as follows: 

\heading{\textbf{Data Confidentiality during Computation}} 
Data sharee and data sharer want to jointly compute a function over their inputs while keeping those inputs private. For example, the data collected from different sensors in the IoT system may be aggregated to generate the targeted result; the cloud and the clients may cooperate to provide appropriate services. At the same time, the private information and secret data should be protected.  The computation procedures and results on BDS should only be known by the data sharer and sharee during and after computation. Unlike traditional cryptographic scenarios, where cryptography ensures security and integrity of communication or storage and the adversary is supposed an outsider from the system of participants (an eavesdropper on the sender and receiver), the cryptography techniques in this model should protect participants' privacy from each other. 
 
\heading{\textbf{Data Confidentiality during Searching}} 
Data sharers want to store data in ciphertext
form while keeping the functionality to search keywords in
the data, i.e, to protect the privacy of data, data sharer may choose to encrypt the data before uploading them
to the cloud. However, while encryption provides confidentiality
to data, it also sacrifices the data sharers’ ability to query a special segment in their data. The search requests and results on big data should only be known by the data sharer during and after searching.

\heading{\textbf{Access control}} Traditionally, cryptography is about providing secure communication over insecure channels, meaning that we want
to protect honest parties from external adversaries: only the party who has the decryption key can learn the
message. In recent applications, more complicated issues have been introduced, i.e, consists in not trusting
everybody with the same information. Access control allows decryption depending on who you are and which keys you have, you can access
different parts of the information sent. In other words, access control requirement deals with the issue that someone should only be able to decrypt a ciphertext if the person holds a key for "matching attributes" where user keys are always issued by some trusted party. 

\heading{\textbf{Delegation rights}} This notion deals with the problem of data sharing between different receipts. We distinguish two types of delegation rights:  \emph{decryption rights delegation} and \emph{signing rights delegation}. In decryption rights delegation, it turns a ciphertext intended for one data user into a ciphertext of the same data intended for another data user without revealing any information about the data or the secret keys. In signing rights delegation, it allows an entity to delegate its signing rights to another.

\begin{figure}[!htbp]
    \centering
    \scalebox{0.65}{\includegraphics[width=0.4\linewidth]{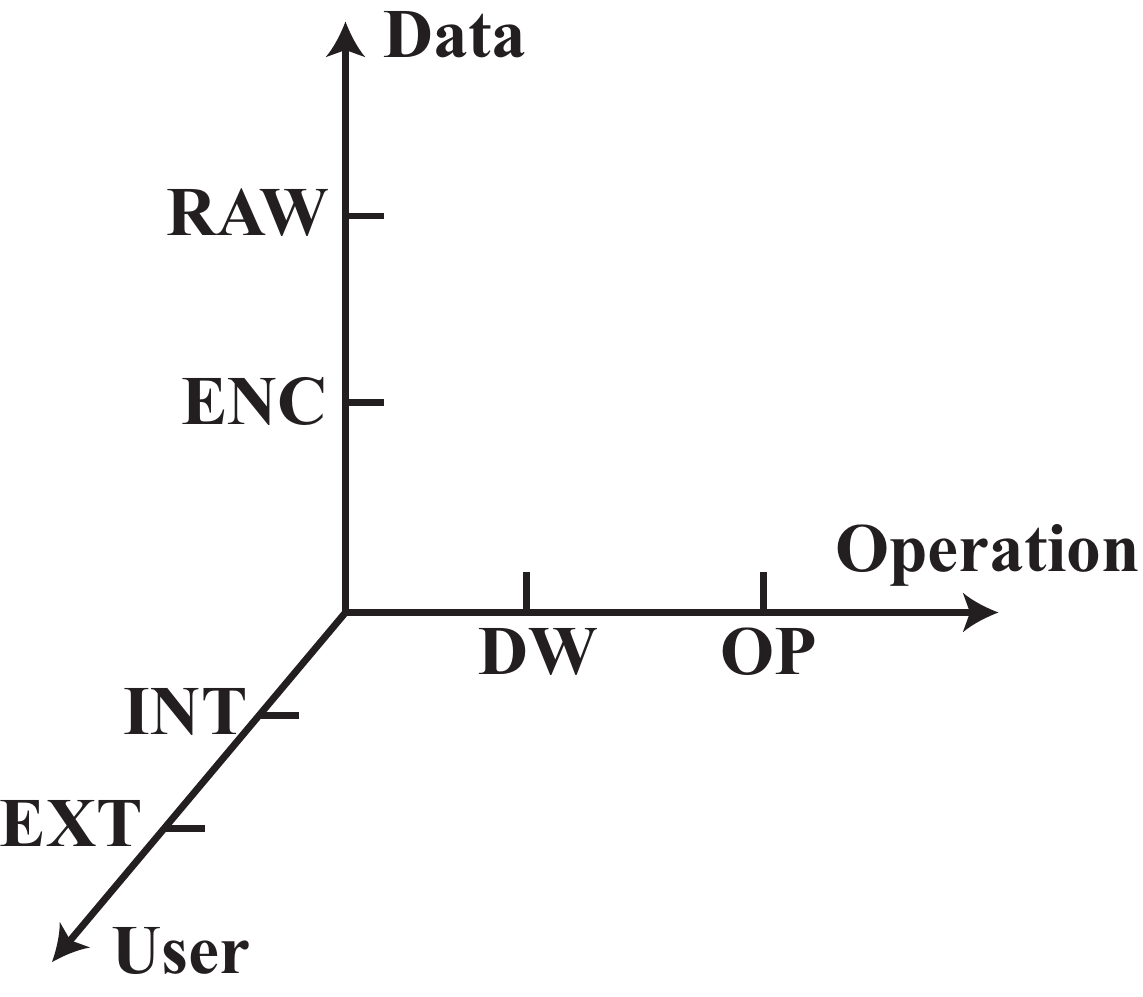}}
    \caption{Assumptions of BDS security}
    \label{fig:big-data-assumption}
\end{figure}
\begin{table*}[tbp]
\centering
\caption{Classification of techniques according to assumptions and fulfilled security requirements}
\label{tab:Summarization security-techniques}
\scalebox{0.68}{\begin{tabular}{p{.18\textwidth} p{.68\textwidth} p{.52\textwidth}} \toprule
\textbf{Assumptions} & \textbf{Fulfilled Security requirements} & \textbf{Security Techniques}
\\ \toprule
\makecell[l]{
Asm.1\\
$\mathsf{INT\mathchar`-OP\mathchar`-ENC}$
}
&
\makecell[l]{
Confidentiality during Computation \\
Confidentiality during Searching \\
Confidentiality during Outsourcing \\
Confidentiality during Outsourcing \& Integrity\\
Availability
}
&
\makecell[l]{
$\mathcal{HE}$  \\
$\mathcal{SSE, PIR}$\\
$\mathcal{ORAM}$ \\
$\mathcal{AE}$\\
Storage path encryption 
}
\\ \midrule
\makecell[l]{
Asm.2\\
$\mathsf{EXT\mathchar`-OP\mathchar`-ENC}$
}
&
\makecell[l]{
Confidentiality during Computation \\
Confidentiality during Searching \\
Confidentiality during Outsourcing\& Integrity\\
Confidentiality during Outsourcing, Integrity \& Non-repudiation\\
Availability
}
&
\makecell[l]{
$\mathcal{FE},\mathcal{HE}$ \\
$\mathcal{PEKS}$ \\
$\mathcal{AE}$, Signcryption \\
Signcryption\\ 
Storage path encryption
}
\\ \midrule
\makecell[l]{
Asm.3\\
$\mathsf{INT\mathchar`-DW\mathchar`-ENC}$
}
&
\makecell[l]{
Confidentiality during Outsourcing\\
Integrity\\
Availability
}
& 
\makecell[l]{
N/A  \\
$\mathcal{DS}$, $\mathcal{MAC}$, $\mathcal{PDP}$\\
Storage path encryption
}
\\ \midrule
\makecell[l]{
Asm.4\\
$\mathsf{EXT\mathchar`-DW\mathchar`-ENC}$
}
&
\makecell[l]{
Confidentiality during Outsourcing\& Rights Delegation \\
Confidentiality during Outsourcing \& Access control\\
Confidentiality during Outsourcing \& Integrity \\
Confidentiality, Integrity \& Non-repudiation \\
Availability
}
&
\makecell[l]{
$\mathcal{PRE}$\\
$\mathcal{ABE}$, $\mathcal{IBE}$,$\mathcal{ABHE}$\\
$\mathcal{AE}$, Signcryption\\
Signcryption\\ 
Storage path encryption
}
\\ \midrule
\makecell[l]{
Asm.5\\
$\mathsf{INT\mathchar`-DW\mathchar`-RAW}$
}
&
\makecell[l]{
 Integrity\\
Availability
}
&
\makecell[l]{
$\mathcal{DS}, \mathcal{MAC},\mathcal{PDP}$\\
Storage path encryption
}
\\ \midrule
\makecell[l]{
Asm.6\\
$\mathsf{EXT\mathchar`-DW\mathchar`-RAW}$
}
&
\makecell[l]{
Integrity\\
Integrity \& Non-repudiation \\
Confidentiality during Outsourcing, Integrity \& Non-repudiation \\
 Integrity \& Rights Delegation  \\
Availability
}
&
\makecell[l]{
$\mathcal{DS}$, $\mathcal{MAC}$, $\mathcal{PDP}$ \\
$\mathcal{DS}$\\
Signcryption\\
$\mathcal{PS}$\\
Storage path encryption
}
\\ \midrule
\makecell[l]{
Asm.7\\
$\mathsf{INT\mathchar`-OP\mathchar`-RAW}$
}
&
\makecell[l]{
Confidentiality during Computation \\
Confidentiality during Searching \\
Integrity  \\
Availability
}
&
\makecell[l]{
$\mathcal{PSI},\mathcal{MPC}$\\
$\mathcal{PIR}$  \\
$\mathcal{VC}$, $\mathcal{DS}$, $\mathcal{MAC}$, $\mathcal{PDP}$\\
Storage path encryption
}
\\ \midrule
\makecell[l]{
Asm.8\\
$\mathsf{EXT\mathchar`-OP\mathchar`-RAW}$
}
&
\makecell[l]{
Confidentiality during  Computation \\
Integrity \\
Integrity \& Non-repudiation\\
Integrity \& Right Delegation \\
Availability
}
&
\makecell[l]{
$\mathcal{PSI},\mathcal{MPC}$ \\
$\mathcal{DS}$, $\mathcal{MAC}$ , $\mathcal{PDP}$\\
$\mathcal{DS}$ \\
$\mathcal{PS}$\\
Storage path encryption
}
\\
\bottomrule 
\end{tabular}}
\end{table*}

\subsection{Security Assumptions}
\label{Security assumptions}
In this section, based on the proposed general model for BDS in Section \ref{SystemModel}, we propose the security assumptions needed in BDS systems according to the three following dimensions: user accessibility, data usage and the form of shared data,  as shown in Fig. \ref{fig:big-data-assumption}. Depending on what action can be performed
by each type of user in BDS introduced in Section \ref{SystemModel}, under which form of data, different cryptographic solutions (or the combination of them) should be used to guarantee the security of BDS operations, we divide the assumptions into the following category: 

\labelText{Assumption 1}{label:text1}
($\mathsf{INT},\mathsf{OP},\mathsf{ENC}$). In such assumption, the encrypted data from the data sharer can only be operated, such as data outsourcing, calculation, searching, viewing. However, it can not be externally accessed, e.g., users store their own data in server and operate/search on it.

\labelText{Assumption 2}{label:text2} ($\mathsf{EXT},\mathsf{OP},\mathsf{ENC}$). In such assumption, the encrypted data from the data sharer can only be operated but the sharee which can be from the outside of system, e.g., access to a server storage, which means that it needs an extra secure channel compared with assumption 1.

\labelText{Assumption 3}{label:text3} ($\mathsf{INT},\mathsf{},\mathsf{ENC}$).  In such assumption, the encrypted shared data can be downloaded and then retrieved by the data owner within the system.

\labelText{Assumption 4}{label:text4} ($\mathsf{EXT},\mathsf{DW},\mathsf{ENC}$).  In such assumption, the encrypted shared data can be downloaded and retrieved by data sharee from the outside of system. Retrieval is a stronger assumption than operation since it delivers all data usage to the data sharee, which means they need additional security requirements such as decryption rights delegation between the data sharer and the data sharee and ensuring the non-repudiation property.

\labelText{Assumption 5}{label:text5} ($\mathsf{INT},\mathsf{DW},\mathsf{RAW}$). In such assumption, data is not encrypted initially and need to be fully downloaded, e.g., the data sharer upload their own data within the server and later download it.

\labelText{Assumption 6}{label:text6} ($\mathsf{EXT},\mathsf{DW},\mathsf{RAW}$). In such assumption, we require that the data sharee can also download the data securely from the server, which means they need extra requirements such as signing rights delegation and ensuring the non-repudiation property.

\labelText{Assumption 7}{label:text7} ($\mathsf{INT},\mathsf{OP},\mathsf{RAW}$). In such assumption, the data sharer operates on his own raw data. E.g. a data owner may need some operations that only work on raw data in server storage. So this also requires the consideration in terms of verifiable outsourced computation integrity from the server.

\labelText{Assumption 8}{label:text8} ($\mathsf{EXT},\mathsf{OP},\mathsf{RAW}$). In such assumption, the data sharee operates on data sharer's raw data. Comparing with Assumption 7, there is extra security requirements in terms of the signing rights delegation and the non-repudiation.

In table \ref{tab:Summarization security-techniques}, we summarize the security techniques detailed in Section \ref{security techniques} that are needed to fulfil the security requirements described in Section \ref{Security requirements} according to each security assumptions described in  \ref{Security assumptions}. 

\subsection{Security Techniques}
\label{security techniques}

In this section, we summarize the existing techniques of BDS that can be leveraged
to enhance the security and privacy of the introduced BDS in Section[\ref{SystemModel}].
For each of the presented techniques, we use the following outline: First, we provide a
high-level overview of what protections the cryptographic techniques provides and how it
can be used in BDS.
Second, we give a more detailed definition of the security achieved and the critical
limits of the technique. Finally, we give an in-depth survey of the literature and state of the art developments for the technique to illustrate the differences between individual
schemes and their potential uses-cases.

\textsc{\textbf{Message Authentication Code ($\mathcal{MAC}$)}} \label{MAC} This is a small piece of information used to authenticate a message. In BDS system, a data sharee needs to be assured that a data comes from a legitimate data sharer (authentication) and not from an attacker. This also includes the assurance that the message was not modified during transmission (integrity). The $\mathcal{MAC}$ algorithm takes a data D and secret key and outputs a $\mathcal{MAC}$ value or "tag". $\mathcal{MAC}$s only use secret keys, and rely on symmetric encryption. However, to function as intended the $\mathcal{MAC}$ must be able to resist plaintext attacks even if a hacker knows the secret key. Although the hacker can create their own $\mathcal{MAC}$s from the key, the $\mathcal{MAC}$ algorithm must be strong enough to make it impossible for the hacker to calculate the $\mathcal{MAC}$ for other messages. $\mathcal{MAC}$s can be built from hash functions; these are known as keyed hash functions. One advantage of $\mathcal{MAC}$s over $\mathcal{DS}$ is that they are much faster than digital signatures since they are based on either block ciphers or hash functions. The algorithms that are more commonly used in modern applications and one specifically designed for constrained platforms: CMAC \cite{CMAC}, PMAC1 \cite{10.1007/3-540-48910-X_9}, GMAC \cite{GCM} and Marvin \cite{Simplcio2009TheMM}. For more details, we refer the interested reader to the survey \cite{article}.

$\mathcal{MAC}$ consists of a tuple of algorithms $(\mathsf{KeyGen}, \mathsf{MAC}, \mathsf{Verify})$ satisfying:
\begin{description}
\item $\mathsf{KeyGen}$ (key-generator) gives the key k on input $1^n$, where n is the security parameter.
\item $\mathsf{MAC}$ (signing) outputs a tag t on the key $\sk$ and the input string x.
\item $\mathsf{Verify}$ verifying) outputs accepted or rejected on inputs: the key $\sk$, the string x and the tag t.
\end{description}

\textsc{\textbf{Digital Signature ($\mathcal{DS}$)}}  \label{DS}  $\mathcal{DS}$
were originally proposed by Diffie and Hellman in \cite{1055638} $\mathcal{DS}$ and Rivest, Shamir, and Adelman \cite{10.1145/359340.359342}.  $\mathcal{DS}$ deals with the problem of data authentication and integrity in the asymmetric (public key) setting. In BDS system, a data sharee needs to be assured that a data comes from a legitimate data sharer (authentication) and not from an attacker. This also includes the assurance that the message was not modified
during transmission (integrity). $\mathcal{MAC}$s solved this problem but for the symmetric-key setting. By opposition to $\mathcal{MAC}$s, digital signatures have the advantage of being \emph{publicly verifiable} and \emph{non-repudiable}. Public verifiability implies the transferability of signatures and, thus, signatures prove useful in many
applications, including BDS systems.

Cramer and Shoup \cite{10.1145/357830.357847} and Gennaro, Halevi, and Rabin \cite{10.1007/3-540-48910-X_9} proposed the first signature schemes that are practical and whose security analysis does not rely on an ideal random function based on the so-called Strong RSA assumption.  

A $\mathcal{DS}$~\cite{Kat10} is a tuple,
$\mathcal{DS}= (\mathsf{KeyGen}, \mathsf{Sign}, \mathsf{Verify})$, of probabilistic polynomial-time
algorithms satisfying:
\begin{description}
\item[$(\sk,\pk) \gets \mathsf{KeyGen}(\secparam)$] On input security parameter $\secparam$,
  key generation algorithm $\mathsf{KeyGen}$ produces a pair $(pk,sk)$ of
  matching public and private keys.
\item[$\sigma \gets \mathsf{Sign}(\sk,m)$] Given a message $m$ in a set $\mathcal{M}$ of messages
  and a private keys $\sk$, signing algorithm $\Sign$ produces a
  signature $\sigma$.
\item[$\{0,1\} \gets \mathsf{Verify}(\pk, m, \sigma)$] Given a signature $\sigma$, a message
  $m \in \mathcal{M}$ and a public key $\pk$, the verifying algorithm $\mathsf{Verify}$
  checks whether $\sigma$ is a valid signature on $m$ with respect to
  $\pk$.
\end{description}

\textsc{\textbf{Signcryption}}  \label{signcryption} In some cases of BDS, we require the confidentiality and authenticity separately, but other cases we require them simultaneously. To achieve this special requirement, Signcryption scheme is used. The first signcryption scheme was introduced by Yuliang Zheng in 1997 \cite{Zheng97} based on  elliptic curve cryptography.  Signcryption is a multi-user primitive that is complex to design \cite{10.1007/3-540-46035-7_6}.

Signcryption Scheme consists of tuple of algorithms defined as follows:

\begin{description}
 \item $(\sk, \pk) \gets \mathsf{Keygen(1^\kappa)}$  is the key generation algorithm which takes a security
parameter $k \in N$ and generates a private/public key pair $(\sk, \pk)$.
 \item $C \gets \mathsf{Signcrypt(1^\kappa, m, \sk_a, \pk_b)}$ takes k, a message m, a sender private key $sk_a$ and
a recipient public key $pk_b$, outputs a ciphertext C. m is drawn from a message space M
which is defined as $\{0, 1\}^n$ where n is some polynomial in k.
\item $(m,\sigma, \pk_a) \gets \mathsf{Unsigncrypt(1^\kappa, C, \sk_b)}$ takes  $\kappa$, C and a private key
$\sk_b$, outputs either a triple $(m,\sigma, \pk_a)$ where $m \in M$, s is a signature and $\pk_a$
is a public key, or reject which indicates the failure of unsigncryption.

\item $\{0,1\} \gets \mathsf{Verify(1^\kappa, m, \sigma, \pk_a)}$ takes k, $m \in M$, a signature $\sigma$ and a public
key $\pk_a$, outputs 1 for a valid signature or 0 otherwise.
\end{description}

\textsc{\textbf{Authenticated Encryption ($\mathcal{AE}$)}} \label{AE} The symmetric analogue of a Signcryption is variously called $\mathcal{AE}$ \cite{Bellare03}, which simultaneously ensure the confidentiality and authenticity of data. However, similarly to $\mathcal{MAC}$, $\mathcal{AE}$ does not provide the non-repudiation. For a comprehensive survey on $\mathcal{AE}$, we refer the interested reader to \cite{cryptoeprint:2019:739}.

Basically, there are three approaches to $\mathcal{AE}$:

Basically, there are three approaches to $\mathcal{AE}$ which are defined as follows:
\begin{description}

\item MAC-then-Encrypt (MtE): We first MAC $m$ under key $\sk_1$ to yield tag $\sigma$ and then encrypt the resulting pair $(m,  \sigma)$ under key $\sk_2$.

\item Encrypt-then-MAC (EtM):  We first encrypt $m$  under key $\sk_2$ to yield ciphertext C and then compute   $\sigma \gets MAC_{\sk_1}(C)$  to yield the pair $(C,  \sigma)$.

\item Encrypt-and-MAC (E\&M):  We first encrypt M under key $\sk_1$ to yield ciphertext C and then compute   $ \sigma \gets MAC_{\sk_1}(m)$  to yield the pair $(C,  \sigma)$.
\end{description}

\textsc{\textbf{Identity Based Encryption ($\mathcal{IBE}$)}} \label{IBE} $\mathcal{IBE}$  \cite{Shamir85} allows an access control that is based on the identity of a data user. To protect the privacy of data stored in the cloud, a data user usually encrypts his data in such a way that certain designated data users can decrypt the data. $\mathcal{IBE}$ is regarded as an alternative to $\mathcal{PKE}$ which is proposed to simplify key management in a certificate based public key infrastructure (PKI) by using human identities like email address or IP address as public keys. To preserve the anonymity of sender and receiver, the first $\mathcal{IBE}$ \cite{Boneh01} scheme was proposed. Several constructions of $\mathcal{IBE}$ have been proposed. From a chronological point of view, the first one is by Boneh and Franklin~\cite{10.5555/646766.704155}, who proposed a straightforward scheme in the random oracle model~\cite{10.1145/168588.168596}. Their construction heavily exploits cryptographic pairings, as well as the power conferred by a random oracle. Later, Boneh and Boyen~\cite{10.1007/978-3-540-28628-8_27} refined this idea and came up with an identity based encryption in the standard model, but relying on what is called a non-static $q$-type assumption, where $q$ is related to the number of queries the adversary makes for key derivation.

An $\mathcal{IBE}$ scheme consists of four algorithms $(\setup, \keygen, \Enc, \Dec)$:
\begin{description}

\item[$\big(\msk, \mpk \big) \gets
    \setup(\secparam)$] The setup algorithm takes the security parameter $\secpar$ and outputs a
    master secret key $\msk$ and a (master) encryption key $\mpk$.

\item[$\ctxt \gets \Enc(\mpk, \ID, \ptxt)$]
    The encryption algorithm  takes
    as input an encryption key $\mpk$, an identifier $\ID$
    and a message $\ptxt \in
    \mathcal{M}_{\secpar}$, and outputs a ciphertext $\ctxt$.

\item[$\sk_{\ID} \gets \keygen(\msk, \ID)$]
    The key derivation algorithm  takes as input an identifier $\ID$
    and outputs the corresponding functional key $\sk_{\ID}$.

\item[$\ptxt \gets \Dec(\sk_{\ID},\ID, \ctxt)$]  The decryption algorithm is a deterministic algorithm that takes as input a key $\sk_{\ID}$ and a ciphertext $\ctxt$ and outputs $\ptxt$ or a special error symbol $\perp$ if decryption fails.

\end{description}

\textsc{\textbf{Attribute Based Encryption ($\mathcal{ABE}$)}} \label{ABE} $\mathcal{ABE}$ technique \cite{Sahai05,Vipul06} is regarded as
the most appropriate technologies used to control big data access in the cloud environment; it allows more secure and flexible as granular access control
is possible and it enables data users to upload their data in encrypted forms to the cloud while sharing them with users possessing certain credentials or attributes.  $\mathcal{ABE}$ can be viewed as a generalization of $\mathcal{IBE}$. An implementation of this scheme describing how this can be used for access control was given by Pirretti et al. \cite{Pirretti}. The class of supported policies was extended to arbitrary boolean formulas by Goyal et al. \cite{10.1145/1180405.1180418} and Bethencourt, Sahai, and
Waters \cite{4223236}.

Basically, $\mathcal{ABE}$ techniques are classified into two major
classes: \emph{Ciphertext-Policy ABE}  ($\mathcal{CP\mathchar`-ABE}$) introduced by Sahai and Waters \cite{Waters11} and \emph{Key-Policy ABE} ($\mathcal{KP\mathchar`-ABE}$) \cite{Nuttapong11}. The $\mathcal{CP\mathchar`-ABE}$ is a
form of $\mathcal{ABE}$  in which keys are associated with attributes and data is encrypted according to
a policy specifying which attributes are needed to decrypt the ciphertext. While in $\mathcal{KP\mathchar`-ABE}$, attributes are always used to describe the access policies and
encrypted data. The user’s secret keys generate using these attributes.
There are some available implementations of $\mathcal{ABE}$. The first efficient implementation
was given by Bethencourt et al. \cite{4223236} using the Pairing-Based Crypto library. Another efficient implementation of $\mathcal{ABE}$ was given by Khoury et al. \cite{6899132}, achieves 3 ms for ABE encryption
and 6 ms for decryption.

$\mathcal{ABE}$  A (Ciphertext-Policy) Attribute Based Encryption ( $\mathcal{CP\mathchar`-ABE}$) scheme consists of four algorithms $\mathsf{ABE}=(\setup, 
\keygen, \Enc, \Dec)$.
\begin{description}

\item[$(\mpk,\msk) \gets \setup(\secparam, \mathcal{U})$]
The setup algorithm takes security parameter and attribute universe
description  $\mathcal{U}$ as input. It outputs the public parameters $\mpk$  and a master key $\msk$.

\item[$CT \gets \Enc(\mpk, M, \mathcal{A})$] The encryption algorithm takes as input the public parameters $\mpk$, a message M, and an access structure $\mathcal{A}$ over the universe of attributes.
The algorithm will encrypt M and produce a ciphertext CT such that only a user
that possesses a set of attributes that satisfies the access structure will be able to
decrypt the message. We will assume that the ciphertext implicitly contains $\mathcal{A}$.

\item[$\sk \gets \keygen(\msk, \mathcal{S})$] The key generation algorithm takes as input the master
key $\msk$ and a set of attributes $\mathcal{S}$ that describe the key. It outputs a private key
$\sk$.

\item[$M \gets \Dec(\mpk, CT, \sk)$] The decryption algorithm takes as input the public parameters $\mpk$, a ciphertext CT, which contains an access policy $\mathcal{A}$, and a private key
SK, which is a private key for a set $\mathcal{S}$ of attributes. If the set S of attributes satisfies
the access structure $\mathcal{A}$ then the algorithm will decrypt the ciphertext and return
a message M.

\end{description}

An $\mathcal{KP\mathchar`-ABE}$ scheme consists of four algorithms:
\begin{description}
\item[$(\mpk,\msk) \gets \setup(\secparam)$] This is a randomized algorithm that takes no input other than the implicit security
parameter. It outputs the public parameters $\mpk$ and a master key $\msk$.
\item[$C \gets \Enc(m,\mathcal{A},\mpk)$] This is a randomized algorithm that takes as input a message m, a set of
attributes $\mathcal{A}$, and the public parameters $\mpk$. It outputs the ciphertext C.
\item[$\sk \gets \keygen(A,\mpk,\msk)$] This is a randomized algorithm that takes as input – an access structure
A, the master key $\msk$ and the public parameters $\mpk$. It outputs a decryption key $\sk$.
\item[$m \gets \Dec(C,\mathcal{A})$] This algorithm takes as input – the ciphertext C that was encrypted under
the set $\mathcal{A}$ of attributes, the decryption key $\sk$  for access control structure A and the public
parameters $\mpk$. It outputs the message m if $\mathcal{A} \in A$.
\end{description}

\textsc{\textbf{Verifiable Computation ($\mathcal{VC}$)}} \label{VC}
 $\mathcal{VC}$ \cite{Babai91} schemes enable a weak data user (internal user) to outsource the computation of a function $\mathcal{F}$ on various inputs to a computationally strong but untrusted cloud, which allows the data user to check
the integrity of the computation.
Most verifiable computation constructions are based on probabilistically checkable proofs (or PCPs)  \cite{Arora98}. A great survey of practical verifiable computation implementations is given in \cite{Walfish15}.
 \labelText{VC}{label:VC} is used which is integrated as follows: A computationally
weak data user outsource the storage of many data items to a computationally strong but untrusted prover (or cloud server). Each data item is labeled with a string $L_i$.
The data user wishes to compute on some subset of D's data a function $f$, and delegates this task to the cloud server.

\begin{figure}[!htbp]
    \centering
    \scalebox{0.70}{\includegraphics[width=.7\linewidth]{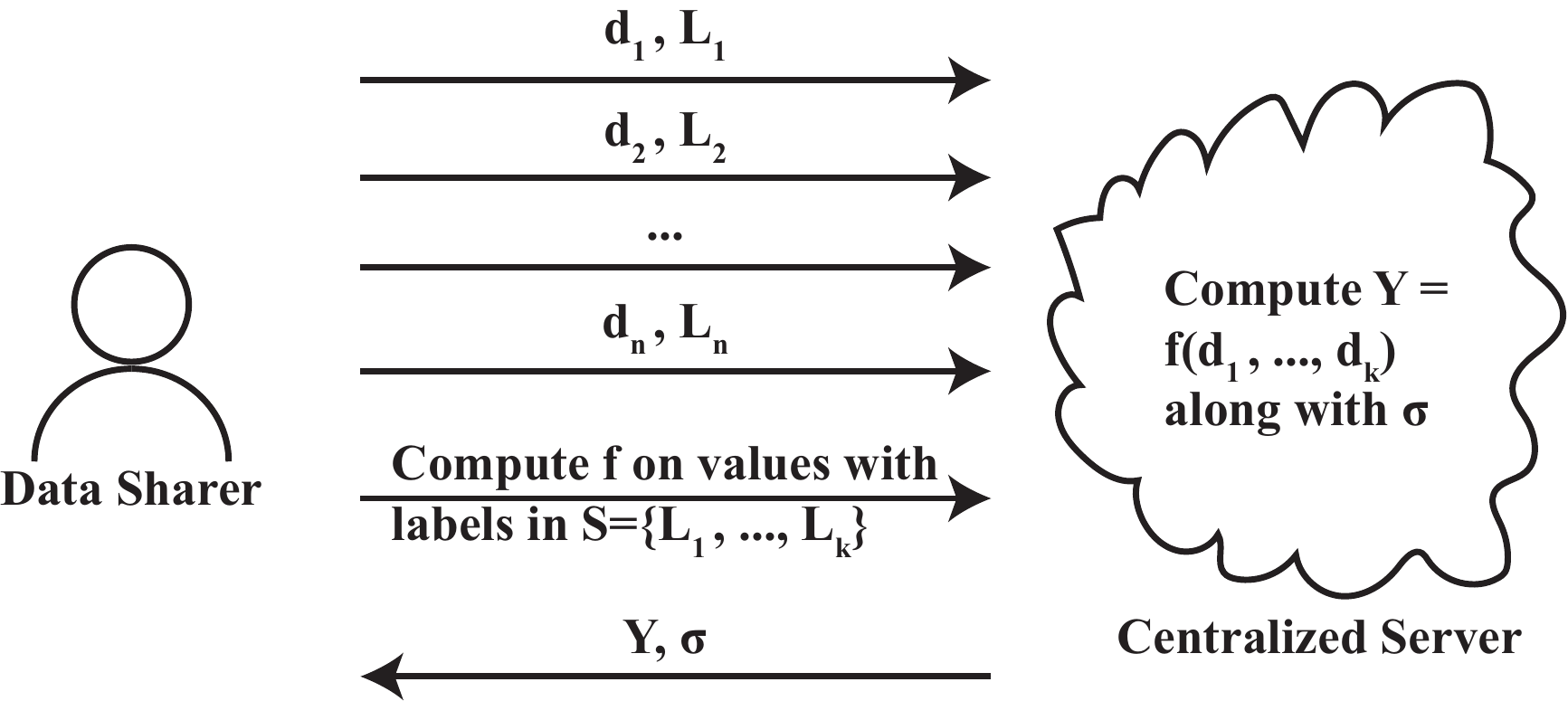}}
    \caption{Verifiable Delegation of Computation on Outsourced Data}
    \label{Verifiable Delegation of Computation on Outsourced Data}
\end{figure}

Given an input: $d_1, \cdots, d_k$ and a function $f$ to evaluate on
$d_1, \cdots, d_k$, the cloud server is expected to produce an output $y$, along with proof $\sigma$ that $y = f(d_1, \cdots, d_k)$ that
the data user can use to confirm the correctness of the computation as shown in Fig. \ref{Verifiable Delegation of Computation on Outsourced Data}.

\textsc{\textbf{Proxy re-encryption ($\mathcal{PRE}$)} } \label{PRE} $\mathcal{PRE}$ provides right delegation and confidentiality of shared data. Encryption scheme such as $\mathcal{IBE}$ and $\mathcal{ABE}$ does not allow the update of ciphertext recipient. Proxy re-encryption ($\mathcal{PRE}$), initially introduced by Blaze\cite{Blaze98}, Bleumer and Strauss \cite{10.1007/BFb0054122} 
in 1998 and improved by Ateniese et al. \cite{Ateniese06,Green06} in 2006,  enables a proxy to transform a encrypted data stored on a cloud storage system under the public key of data user (or delegator) into another encrypted data under the public key of data user (or delegatee) without leaking the underlying encrypted data or private keys of delegator/delegatee to the proxy. This form of public key encryption is the best candidate to ensure the security of sharing data in cloud computing.    $\mathcal{PRE}$ could be combined with $\mathcal{IBE}$, which yields Identity-Based Proxy Re-Encryption $\mathcal{IBPRE}$, where ciphertexts are transformed from one identity to another \cite{Green07}. \labelText{Proxy re$\mathchar`-$encryption}{label:PRE} consists of the following algorithms:

 \begin{description}
\item[ $(\pk,\sk) \gets \KeyGen(\secparam)$] On input the security parameter $\secparam$, the key generation algorithm $\KeyGen$ outputs a key pair (pk,sk).
\item[$rk_{pk \rightarrow pk'} \gets \ReKeyGen(\pk,\sk, \pk'
)$] On input a private key $\sk$ of a delegator
and a public key of a delegatee $\pk$
, algorithm $\ReKeyGen$ outputs a unidirectional re-encryption key $rk_{pk \rightarrow pk'}$.
\item[ $c \sample \Enc(\pk, m)$] On input a public key pk and a message m, algorithm $\Enc$
outputs a ciphertext c.
\item[$c' \gets \ReEnc(rk_{pk \rightarrow pk'} , c)$]On input a re-encryption key $rk_{pk \rightarrow pk'}$ and a ciphertext c'
, algorithm $ \ReEnc$  outputs a ciphertext c' decryptable under the secret
key $\sk'$.
\item[$m \gets \Dec(\sk, \pk, c)$]On input a secret key $\sk$, a public key pk and a ciphertext c', algorithm Dec outputs a message m or the error symbol $\perp$.

 \end{description}

\textsc{\textbf{Proxy Re-Signature}} \label{PS} Proxy re-signatures \cite{10.1007/BFb0054122} should not be confused with the proxy signatures \cite{10.1145/238168.238185,Ivan03proxycryptography}
which definition is given in Dodis et al. \cite{Ivan03proxycryptography}. In
their general construction, Bob’s signature is considererd as a double signature which includes a signature
from Alice and one from the proxy. There is clearly no key transformation from valid Alice's singing secret key 
into Bob's ones. In a proxy re-signature scheme, a semi-trusted proxy is given some information which allows it to
transform data user’s signature on a data  D into data user’s on data D. 

A proxy re-signature scheme is a tuple of (possibly probabilistic) polynomial time algorithms $\mathsf{PS}=(\KeyGen, \mathsf{ReKey}, \mathsf{Sign},\mathsf{ReSign}, \Verify)$, where:
$(\KeyGen,  \mathsf{Sign}, \Verify)$ form the standard key generation, signing, and verification of $\mathsf{DS}$ algorithms as in DS.
\begin{description}

    \item[$rk_{A \rightarrow A} \gets \mathsf{ReKey}$]  On input $(\pk_A, \sk *
A, \pk_B, \sk_B)$, the re-encryption key generation algorithm, ReKey, outputs a
key $rk_{A \rightarrow A}$ for the proxy. (Note: $rk_{A \rightarrow A}$  allows to transform A's signatures into B's signatures
– thus B is the delegator.) The input marked with a  ‘*’ is optional.
 \item[$\sigma_B(m) \gets \mathsf{ReSign}$] On input $rk_{A \rightarrow A}$, a public key $\pk_A$, a signature $\sigma$, and a message m, the re-signature function,
$\mathsf{ReSign}$, outputs $\sigma_B(m)$ if $\Verify(pk_A, m, \sigma)$ and $\perp$ otherwise.

\end{description}

\textsc{\textbf{Searchable Symmetric Encryption ($\mathcal{SSE}$)}} \label{SSE}
$\mathcal{SSE}$ \cite{Curtmola06} aims to provide confidentiality and searchability simultaneously. The data user can delegate a token for a specific query, which allows the server to perform the query over encrypted data.
   One approach to provisioning symmetric encryption with search capabilities is with a so-called 
\emph{secure index} as shown by Goh in\cite{cryptoeprint:2003:216}. The client indexes and encrypts its data collection and sends the secure index
together with the encrypted data to the server. To search for a keyword w, the data user generates and
sends a trapdoor for w  which the server uses to run the search operation and recover pointers to the
appropriate (encrypted) datas \cite{10.5555/2590701.2590705}.
    Alternatively, using oblivious RAMs techniques symmetric searchable encryption can be 
    can be achieved in its full generality (e.g., conjunctions or disjunctions of keywords) with optimal security using the work of Ostrovsky and Goldreich on oblivious RAMs \cite{10.1145/100216.100289,10.1145/233551.233553}. We assume that the client processes the data collection $\mathsf{D}=(\mathsf{D}_1, \cdots, \mathsf{D}_n)$ and sets up a "database" DB that maps every keyword $\mathsf{w}$ in the collection to the identifiers of the documents that contain it. Recall that in our context, we use the term database loosely to refer to a data structure optimized for keyword search (i.e., a search structure). For a keyword $\mathsf{w}$, we'll write $\mathsf{DB}[w]$ to refer to the list of identifiers of documents that contain $\mathsf{w}$. A non-interactive and response-revealing \labelText{SSE}{label:SSE} scheme $(\Setup,\mathsf{Token},\mathsf{Search})$ consists of:

\begin{description}

\item ($\sk,\mathsf{EDB}) \gets \Setup(\secparam,DB)$ a Setup algorithm run by the client that takes as input a security parameter $\secparam$ and a database DB; it returns a secret key $\sk$ and an encrypted database EDB;
\item $\mathsf{tk} \gets \mathsf{Token}(\sk,\mathsf{w})$ a Token algorithm also run by the client that takes as input a secret key $\sk$ and a keyword $\mathsf{w}$; it returns a token tk;
\item $\mathsf{DB}[w] \gets \mathsf{Search}(\mathsf{EDB},\mathsf{tk})$ a Search algorithm run by the server that takes as input an encrypted database $\mathsf{EDB}$ and a token $\mathsf{tk}$; it returns a set of identifiers $\mathsf{DB}[w]$.
\end{description}

\textsc{\textbf{Public Key Encryption with Keyword Search ($\mathcal{PEKS}$)}} \label{PEKS}  $\mathcal{PEKS}$ was introduced in 2004 by Boneh et al. \cite{10.1007/978-3-540-24676-3_30}, mainly based on public
key encryption algorithms, enables a data sharee to retrieve encrypted data containing some specific keyword from the centralized server.

A data sharer encrypts both their data and index with the public key and uploads
to the remote server provider. A data sharee who has received the corresponding private key can perform
the search operation. He generates the trapdoor he wants to search the keyword with the private key
and sends it to the server. After receiving trapdoor, the server provider enable test whether a given
ciphertext contains the search keyword without knowing the corresponding plaintext of the encrypted data
and the keyword. Then, the server provider returns the query results to the data sharee. Finally, the
sharee can decrypt the encrypted data sent by the server.

A $\mathcal{PEKS}$  scheme consists of
the following algorithms: $\mathsf{(KeyGen,PEKS,Trapdoor,Test)}$:

\begin{description}
    \item[$(\pk_R, \sk_R) \gets \mathsf{KeyGen}(s)$] takes a security parameter, s, and generates a public/private key pair $pk_R$, $sk_R$.
    \item[$(\pk_R,W) \gets \mathsf{PEKS}(\pk_R, \mathcal{W})$]: for a public key $\pk_R$ and a word $\mathcal{W}$, produces a searchable encryption of $\mathcal{W}$.
    \item[$\mathcal{T}_W \gets \mathsf{Trapdoor}(\sk_R, \mathcal{W})$] given Receiver’s private key and a word W produces a trapdoor $\mathcal{T}_W$.
    \item[$(0,1) \gets \mathsf{Test}(\pk_R, S, \mathcal{T}_W )$] given Receiver’s public key, a searchable encryption $S = \mathsf{PEKS}(pk_R, \mathcal{W}_0)$, and
a trapdoor $\mathcal{T}_W =\mathsf{Trapdoor}(\sk_R, \mathcal{W})$, outputs 1 if $\mathcal{W} = \mathcal{W}_0$ and 0 otherwise.
\end{description}

\begin{figure}[!htbp]
            \centering
            \includegraphics[width=.7\linewidth]{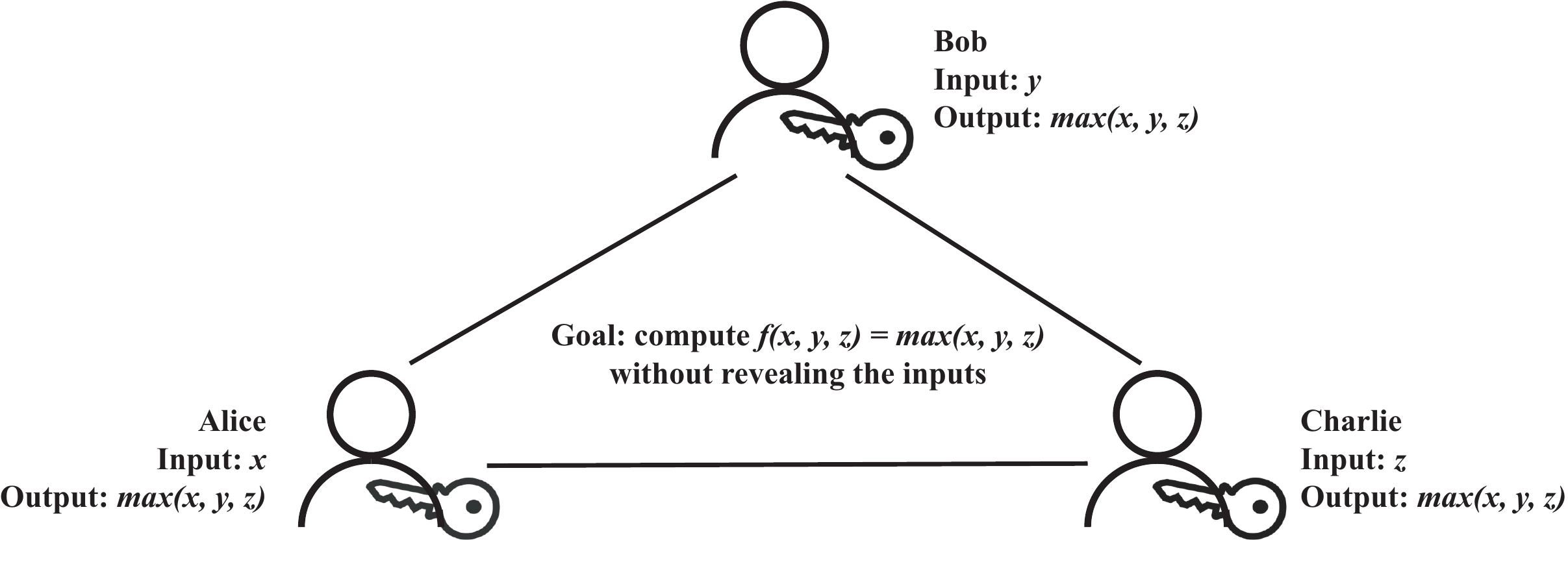}
            \caption{MPC protocol}
            \label{MPC protocol}
        \end{figure}

\textsc{\textbf{Secure Multi Party Compution ($\mathcal{SMPC}$)}} \label{MPC} $\mathcal{SMPC}$ introduced by Yao in 1982 \cite{Yao1982}, is a ‘toolbox’ of cryptographic techniques that allows several different data sharers to jointly analyze data, just as if they have a shared database without violating their underlying sensitive data privacy and only the output of the analysis will be revealed. The concept of MPC is presented a story about two millionaire’s problem to lead to $\mathcal{SMPC}$ processing: two millionaires want to know who is the wealthiest one, while they don’t want to reveal individual wealth to another. However, only some simple functions can be carried out, and complex functions are very demanding in terms of efficiency. Yao \cite{Yao1982} presented the first two-party
protocol for computing functions represented as boolean circuits using a technique called
garbled circuits. Later, Goldreich, Micali, and
Wigderson \cite{10.1145/28395.28420} made two contributions: First, they introduced the first multi-party protocol, also for boolean circuits, with computational security against a semi-honest adversary, and second a general compiler for transforming any protocol with semi-honest security to one with malicious security.
 Ben-Or et al and Chaum et al in \cite{10.1145/62212.62213,10.1145/62212.62214}.       
More formally, $\mathcal{SMPC}$ is stated as follows: Given number of data users: $\mathcal{O}_1, \cdots, \mathcal{O}_n$, each have private data, respectively $\mathcal{D}_1, \cdots,\mathcal{D}_n$. Data users want to compute jointly a public function $\mathcal{F}$ on their private data: $\mathcal{F}(\mathcal{D}_1, \cdots,\mathcal{D}_n)$ while keeping that private data secret. For example, suppose we have three data users Alice, Bob and Charlie, with respective inputs x, y and z denoting their own personal wealth. They want to find out the wealthiest, without revealing to each other how much each of them has as shown in Fig. \ref{MPC protocol}. Mathematically, this translates to them computing:
        
        \[ \mathcal{F}(x,y,z) = max(x,y,z) \]

\begin{figure}[!htbp]
\centering
\scalebox{0.70}{  \includegraphics[width=.7\linewidth]{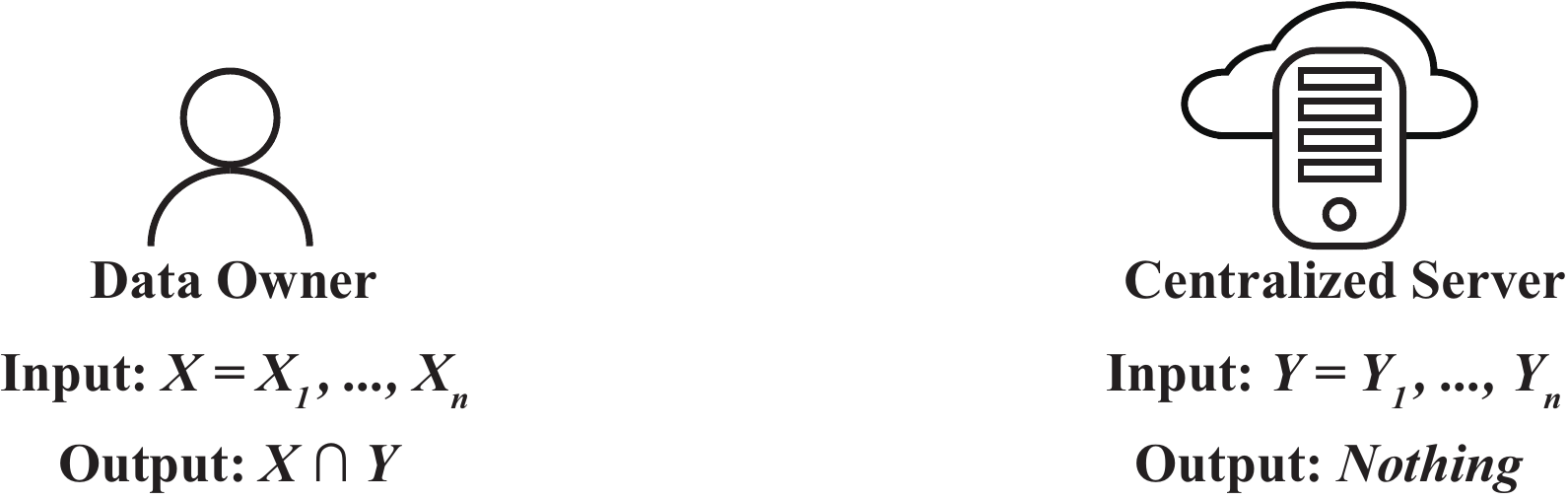}}
\caption{Server-client PSI}
    \label{fig:PSI}
\end{figure}

\textsc{\textbf{Private Set Intersection ($\mathcal{PSI}$)}}  \label{PSI} $\mathcal{PSI}$ \cite{Chen17}  is a powerful tool from $\mathcal{SMPC}$ cryptographic technique that allows two data sharers holding sets, to compare encrypted versions of these sets in order to compute the intersection. For example, $\mathcal{PSI}$ allows to test whether the parties share a common datapoint (such as a location, ID, etc). Many $\mathcal{PSI}$ protocols have been proposed. For example, in \cite{Changyu13}they proposed a protocol based on a novel two-party computation approach which gives better a reasonable efficiency and scalability. 
Among the first protocols for $\mathcal{PSI}$ was \cite{10.1007/978-3-540-24676-3_1} which is based on Oblivious Polynomial
evaluation (OPE), however it requires heavy cost in terms of computational complexity. Later, $\mathcal{PSI}$protocol with reasonable linear computation and communication complexity was introduced in \cite{6234849} by using the Diffie-Hellman protocol (DH).
The most recent and most efficient $\mathcal{PSI}$ protocols are based on either using efficient OT extension and
garbled Bloom filters or hashing to bins. Existing $\mathcal{PSI}$ protocols are compared in \cite{10.1145/3154794}. 
 \labelText{PSI}{label:PSI} Suppose we have two parties: Alice and Bob such that Alice has a set of items:  $\mathcal{A}=(a_1,\cdots ,a_n)$ and Bob has another set:  $\mathcal{B}=(b_1,\cdots ,b_n)$.The goal of $\mathcal{PSI}$ is to allow Alice and Bob to obtain the result of the intersection $\mathcal{A}\cap \mathcal{B}$, under the following privacy restriction: The protocol must not reveal anything about items without revealing any additional information beyond the intersection itself. The server-client $\mathcal{PSI}$ variant is user, the $\mathsf{EXT}$ user will learn the intersection of his encrypted set with the set of the cloud, without the cloud learning intersection of his set with the client as shown in Fig. \ref{fig:PSI}.

\textsc{\textbf{Storage path encryption}}  \label{storage path encryption} Recently Cheng \emph{et al.} \cite{Cheng15}proposed a scheme for secure storage of big data on clouds. In the proposed scheme, the big data are first separated
into many sequenced parts and then each part is stored on
a different storage media owned by different cloud storage
providers. To access to the data, different parts are first collected
together from different data centres and then restored into
original form before it is presented to the data user. 
\labelText{Storage path encryption}{label:Storage path encryption} Data will be first separated into a sequence of $n$ parts and then each part will be stored at $m$ different storage providers. To retrieve the data, different parts are first collected together from different data centres and then restored into original data before sending it to the data user. 

\textsc{\textbf{Oblivious RAM ($\mathcal{ORAM}$)}}  \label{ORAM} This technique was introduced by Goldreich and Ostrovsky \cite{Ostrovsky90}, which enables a data sharer to store data on a cloud server and read/write to individuals locations of the data while hiding the \emph{access pattern}. The security of $\mathcal{ORAM}$ is based on the fact that for any two values $\mathsf{M_1,M_2}$ and equal-size sequences of read/write operations: $\mathsf{S_1,S_2}$, the server cannot distinguish between $\mathcal{ORAM}$ execution with ($\mathsf{M_1,S_1}$)  and ($\mathsf{M_2,S_2}$).
   \labelText{ORAM}{label:ORAM} A client C wants to perform read and write operations on a large database residing on a remote, untrusted server S. The database is encrypted with a symmetric key owned by C. Whenever C wants to perform an operation on the database, it does the following:
  C sends a request to S to download the whole database.
Then, C decrypts the whole database, performs the operation on the desired element, then re-encrypts the database (with the same key). Finally, C re-uploads the re-encrypted whole database to S.

\textsc{\textbf{Proof of Data Possession ($\mathcal{PDP}$)}} \label{PDP} $\mathcal{PDP}$ \cite{10.1145/1315245.1315318} is a cryptographic protocol the provide data integrity verification in remote untrusted servers. Client periodically
 challenges the server to ask relevant evidence that can prove the data exists, then client will compare the relevant with local evidence to verify the integrity of data. $\mathcal{PDP}$ supports probabilistic proof, which means client does not need to challenge and compare all evidence corresponding to the data. This property significantly reduces the computation and communication cost during the protocol procedure. Other variants of $\mathcal{PDP}$ \cite{ateniese2008scalable}  \cite{curtmola2008mr} \cite{erway2015dynamic} are also introduced later for improving efficiency, scalabiltiy, supporting multiple-replicas and file updated. 
 \labelText{PDP}{label:PDP} 

Consider we have a client who owns the data $m$ and wants to store the data on server. The definition of general $\mathcal{PDP}$ as follows:
\begin{description}

\item $(\pk, \sk) \gets \KeyGen(\secparam)$ is a probabilistic key generation algorithm run by the client to setup the scheme. It takes a security parameter $k$ as input, and returns a key pair of public and secret key $(\pk, \sk)$.
$T_{m} \gets \mathsf{TagBlock} (\pk, \sk, m)$ is a algorithm run by the client to generate the verification metadata. It takes as inputs a public key $\pk$, a secret key $\sk$ and a file block $m$, and returns the verification metadata $T_{m}$.

\item $\mathcal{V} \gets \mathsf{GenProof}(\pk, F, c, \Sigma)$ is run by the server to generate a proof of data possession. It takes a public key $\pk$ as input, an ordered collection $F$ of blocks, a challenge $c$ and an ordered collection $\Sigma$ which is the verification metadata corresponding to the blocks in $F$. It returns a proof of data possession $\mathcal{V}$ for the blocks in $F$ that are determined by the challenge $c$. 

 \item $\{ 1, 0\} \gets \mathsf{CheckProof}(\pk, \sk, c, \mathcal{V})$  is run by the client to verify the proof of possession. It takes as inputs a public key $\pk$, a secret key $\sk$, a challenge $c$ and a proof of possession $\mathcal{V}$. It returns whether $\mathcal{V}$ is a correct proof of possession for the blocks.
\end{description}

\heading{\textbf{Homomorphic Encryption ($\mathcal{HE}$)}}  \label{HE} Due to cost-efficiency reasons, data sharers are usually outsourcing their own data to server which can provide access to the data as a service, $\mathcal{HE}$ which introduced firstly by Rivest et al. \cite{Rivest1978} whose main benefit is that for some operations can be allowed over the data user's encrypted data without decrypting it, which allows to produce result that is still encrypted but when decrypted by the user it matches exactly the result that would be obtained if the same computational operations had been performed on the user's raw data as opposed to the uploaded enrypted data. 

One of the most common scenarios where $\mathcal{HE}$ can be used is in outsourced computations: a data sharer sends encrypted data to a server and asks this latter to evaluate a function $\mathcal{F}$ on this encrypted data. The inputs and outputs of the computation are encrypted with the client’s secret/public key and the server manipulates only encrypted data.

$\mathcal{HE}$ schemes can be roughly classified into 3 following types:

 \begin{itemize}
     
    \item Somewhat Homomorphic Encryption ($\mathcal{SHE}$): In $\mathcal{SHE}$ scheme, both addition and multiplication operation is allowed but with only a limited number of times.

\item Fully Homomorphic Encryption ($\mathcal{FHE}$):  In $\mathcal{FHE}$ scheme, allow to do unlimited number of homomorphic operations by bootstrapping \cite{Gentry09}. 
     
\item Partially Homomorphic Encryption ($\mathcal{PHE}$): In $\mathcal{PHE}$ scheme, are in general more efficient than $\mathcal{SHE}$ and $\mathcal{FHE}$, mainly because only one type of homomorphic operation is allowed on the encrypted message, i.e., either addition or multiplication operation, with unlimited number of times.
 \end{itemize}
  Early homomorphic cryptosystems such as RSA \cite{Rivest1978}, El Gamal \cite{10.1007/3-540-39568-7_2}, and Paillier \cite{10.1007/3-540-48910-X_16} can only support
a single operation on ciphertexts such as addition, multiplication, or XOR and are called partially
homomorphic. New cryptographic solutions for computation outsourcing became possible after
Gentry’s discovery of the first viable $\mathcal{FHE}$, which solved a long-standing major problem in cryptography and theoretical computer science \cite{Gentry09}.
 The first plausible and achievable  $\mathcal{FHE}$ scheme, was introduced in 2009 in the seminal work of Gentry \cite{10.5555/1834954}, which which allows any computable function (both additions and multiplications) to perform on the encrypted data. Gentry split the $\mathcal{FHE}$ problem into two components: the design of a somewhat homomorphic encryption scheme ($\mathcal{SWHE}$) that allows a
limited number of $\textsc{Eval}$ operations, which allows fully homomorphic encryption using a boostrapping algorithm and the multiple application of $\mathcal{SWHE}$. The initial scheme was implemented by Gentry and Halevi \cite{10.1007/978-3-642-20465-4_9}.
  
Unlike the public key encryption, which has three security procedures, i.e., key generation, encryption and decryption; there are four procedures in $\mathcal{HE}$ scheme, including the evaluation algorithm.

The notation $\mathcal{E}(x)$ is used to denote the encryption of the message x.

Let $\mathcal{E}(m_1)=m_1^e$ and $\mathcal{E}(m_2)=m_2^e$. 

Then, Addition Homomorphism: 
\[ \mathcal{E}(m_1) + \mathcal{E}(m_2) =m_1^e+m_2^e= (m_1+m_2)^e=\mathcal{E}(m_1+m_2) \]

Multiplication Homomorphism: \[ \mathcal{E}(m_1)\cdot \mathcal{E}(m_2) =m_1^e \cdot m_2^e= (m_1\cdot m_2)^e=\mathcal{E}(m_1 \cdot m_2) \]
 
More formally,  Let the message space  $(\mathcal{M},  \mathsf{o})$  be a finite (semi-)group, and let $k$ be  the  security parameter a \labelText{HE}{label:HE} on  $\mathcal{M}$ is a quadruple $(\mathsf{KeyGen}, \mathsf{Enc}, \mathsf{Dec}, A)$ of probabilistic, expected polynomial time algorithms, satisfying the following functionalities:

\begin{description}
\item $\mathsf{k} \gets \mathsf{KeyGen}(\secparam)$] On input $\secparam$ the Key Generation algorithm $\mathsf{KeyGen}$ outputs an encryption/decryption key pair $\mathsf{k}=(\mathsf{k}_e, \mathsf{k}_d)=k \in \mathcal{K}$, where $\mathcal{K}$ denotes the key space.

\item $c \gets \mathsf{Enc}(\secparam,\mathsf{k}_e,m)$ On inputs $\secparam$, $\mathsf{k}_e$, and an element $m \in M$  the encryption algorithm E outputs a ciphertext $c \in C$,  where C denotes the ciphertext space.

\item $m \gets \mathsf{Dec}(c,\secparam,\mathsf{k})$] The decryption algorithm $\mathsf{Dec}$ is deterministic. On inputs security parameter,$\mathsf{k}$ and an element $c \in C$  it outputs an element in the message space $\mathcal{M}$ so that for all $m \in \mathcal{M}$ it holds:  if $c=\mathsf{Enc}(\secparam,\mathsf{k}_e, m)$  then $\textsc{Pr}[\mathsf{Dec}(\secparam,\mathsf{k},c) \ne m]$ is negligible, i.e., it holds that $\textsc{Pr}[\mathsf{Dec}(\secparam,\mathsf{k},c) \ne m] < 2-k$ .
\end{description}
Homomorphic Property: A is an algorithm that on inputs $\secparam,\mathsf{k}_e$, and elements $(c_1,c_2) \in C$ outputs an element $c_3 \in C$ so that for all $m_1, m_2 \in \mathcal{M}$  it holds: if $m_3=m_1 ~\mathsf{o}~ m_2$ and $c_1=\mathsf{Enc}(\secparam,k_e, m_1)$, and $c_2=\mathsf{Enc}(\secparam, k_e, m_2)$, then $\textsc{Pr}[\mathsf{Dec}(A(\secparam, k_e, c_1, c_2))] \ne m_3]$ is negligible.

The security of the most practical $\mathcal{HE}$ schemes is based on the Ring-Learning With Errors ($\mathsf{RLWE}$) problem, which is a hard mathematical problem related to high-dimensional lattices. Namely, the security assumption of these encryption schemes states that if the scheme can be broken efficiently, then the $\mathsf{RLWE}$ problem can be solved efficiently. 

\heading{\textbf{Functionnal Encryption ($\mathcal{FE}$)}}  \label{FE} $\mathcal{FE}$  is a public key construction, on which it is possible to produce functional secret keys allowing a party to evaluate a specific function $\mathcal{F}$ (generally public) on an encrypted input during its decryption. So the input is encrypted and the output is in cleartext: the party performing the (functional) decryption learns the result of the function on the specific data, but nothing else. A $\mathcal{FE}$ scheme enables a data sharer to encrypt a data set and share the ciphertexts to a data sharee such that the data sharee can obtain a (specific) function value of the data set from the ciphertexts, but nothing more about the data set itself. $\mathcal{FE}$ has been properly formalized in 2011 by Boneh, Sahai and Waters in \cite{10.1007/978-3-642-19571-6_16}. Nowadays, there are no known $\mathcal{FE}$ schemes that can be used to efficiently evaluate general functions. However, the literature proposes multiple efficient constructions to evaluate linear and quadratic functions.

 A $\mathcal{FE}$ for a functionality $\mathcal{F}$ defined over a tuple of four PPT algorithms $(\setup, \keygen, \Enc, \Dec)$ defined as follows:
\begin{description}
\item $(\mpk, \msk) \gets \setup(\secparam)$ On input $\secparam$ generate a public $\mpk$ and master secret $\msk$ key pair.
\item $\sk \gets \keygen(\mpk, k)$ On input  $\mpk$ and k generate secret key for k.
\item $c \gets \Enc(\mpk, x)$ Encrypt message x.
\item $y \gets \Dec(\msk, c)$ On inputs $\msk$ to compute $\mathcal{F}(k, x)$ from c.
\end{description}

\heading{\textbf{Attribute-Based Homomorphic Encryption ($\mathcal{ABHE}$)}}  \label{ABHE} The first $\mathcal{ABHE}$ was constructed by Gentry et al. in \cite{10.1007/978-3-642-40041-4_5} from the Learning with Errors (LWE) problem. 

$\mathcal{ABHE}$ a data sharer describes a policy (predicates) while
encrypting his data, and a trusted party issues a decryption key for
the attributes and distributes them among data sharees. A data sharee can decrypt this ciphertext if his attributes satisfy data user’s defined
policy.

The only known way to achieve fully $\mathcal{ABHE}$ (i.e. where all circuits can be evaluated) is through indistinguishability obfuscation \cite{6686139}, especially the construction in \cite{10.1007/978-3-319-12280-9_1}.

An $\mathcal{ABHE}$ scheme is a tuple of probabilistic polynomial time (PPT) algorithms $\mathsf{ABHE}=(\mathsf{Setup},\mathsf{KeyGen},\mathsf{Encrypt},\mathsf{Decrpt},\mathsf{Eval})$  where   $\mathsf{Setup}$,  $\mathsf{KeyGen}$,   $\mathsf{Encrypt}$  are defined equivalently to $\mathcal{KP\mathchar`-ABE}$[\ref{ABE}]. $m \gets \mathsf{Dec}(k_{f_1},\cdots,k_{f_k},c)$]: On input a sequence of $k \leq \mathcal{K}$ secret keys for policies $f_1, \cdots, f_k \in \mathcal{F}$ and a ciphertext c, output a plaintext $m\in \mathcal{M}$   iff every attribute associated with c is satisfied by at least one of the   $f_i$. Otherwise output $\perp$. $c' \gets \mathsf{Eval}(\mathsf{pp},C,c_1,\cdots ,c_l)$ ]: On input public parameters   $\mathsf{pp}$ , a circuit   $C\in \mathcal{C}$  and ciphertexts  $c_1,\cdots ,c_l \in \mathcal{C}$, output an evaluated ciphertext  $C'\in \mathcal{C}$.

\heading{\textbf{Private Information Retrieval (PIR)}} $\mathcal{PIR}$ is a protocol that allows a user to retrieve an item from a server in possession of a database without revealing which item is retrieved, wich is very useful in the cloud outsourcing context. $\mathcal{PIR}$ is a weaker version of 1-out-of-n oblivious transfer, where it is also required that the user should not get information about other database items. An important goal in $\mathcal{PIR}$ works is to reduce the amount of communication required between the server and the querier. Chor et al. \cite{10.1145/293347.293350} introduced the first $\mathcal{PIR}$  scheme in an information-theoretic model with multiple non-colluding
servers. Shortly thereafter, several works provided $\mathcal{PIR}$  under the assumption that certain
cryptographic problems are hard \cite{10.1007/11523468_65,10.1007/3-540-48910-X_28,10.1007/3-540-45539-6_9}.

More formally, a $\mathcal{PIR}$  is an interactive protocol between two parties: a database $\mathcal{D}$ and a user $\mathcal{U}$. The
database stores a data string $\{0,1\}^n$, and the user has an index $i \in [n]$. In the first round , the protocol does the following (a) the user send a query to the database, this is generated by
an efficient randomized query algorithm, taking as an input the index i and a random string
$r_U$; (b) The database sends an an answer to the user; this is generated by an efficient deterministic answer algorithm, taking as an input the query sent by the user and the
database x; and (c) The user applies an efficient reconstruction function  (taking as an
input the index i, the random string $r_U$, and the answer sent by the database). 

\section{Privacy in BDS}
\label{sec:privacy}

In recent years, with the increase in data demand and
the development of significant data sharing and the CIA’s
requirements, the concept of Privacy has become a new indispensable requirement.
The privacy concerns discussed in this article are mainly divided
into two aspects: data privacy and user privacy which are defined in Section ~\ref{Introduction}.

In this section, first, we discuss the data privacy requirements of BDS, each of such requirements is
targeted at one type of known vulnerabilities. Second, we describe the basic (and inherent) privacy-preserving
techniques. Finally, we discuss techniques to achieve user privacy, which is desired by many BDS applications.

\subsection{Data Privacy requirements in BDS.}
\label{Privacy}

Data Privacy is one of the most important concept of our time, yet it is also one of the most very hard to achieve.  Privacy Preserving Data Publishing (PPDP), is the process through which we provide methods and tools for publishing useful information while preserving data privacy. Recently, PPDP has has been given a considerable object of attention among researchers \cite{10.1145/1749603.1749605,9298747,rajendran2017study}. To achieve PPDP, first the data sharer collects data from individuals.  Then, the data sharer prepares the data to be processed and anonymized. Finally, the processed/anonymized
data is sent to the data sharee for further analysis or research purposes.  The original data are assumed
to be sensitive and private if it contains the following four attributes:

\begin{itemize}
\item Identifier (ID): The attributes which can be used to
uniquely identify a person e.g., name, driving license
number, and mobile number etc.
\item Quasi-identifier (QID): The attributes that cannot
uniquely identify a record by themselves but if linked
with some external dataset may be able to re-identify the
records.
\item Sensitive attribute (SA): The attributes that a person may
want to conceal e.g., salary and disease.
\item Non-sensitive attribute (NSA): Non-sensitive attributes
are attributes which if disclosed will not violate the
privacy of the user. All attributes other than identifier,
quasi-identifier and sensitive attributes are classified as
non-sensitive attributes.
\end{itemize}
Below, we discuss the three types of data privacy requirements in BDS, which we define as follows: 

\textbf{Data Privacy during Downloading.}\label{data publishing} 
It deals with what someone does once he have obtained the data containing sensitive information.
Actions include the following: \circled{1} Aggregation: where data is combined, \circled{2}  identification: where
data is connected to an individual. \circled{3}  secondary use: where data is used for a reason other than was
intended \circled{4} exclusion: where data isn't revealed to the person it was collected from. \circled{5} insecurity:
where data is leaked. The publishing procedures on big data should not contain sensitive information of individuals.

\textbf{Data Privacy during Computation.}\label{data computation} Some operations can be run over encrypted data, e.g. a number of databases around the world currently host a wealth of genomic data that is very useful to researchers conducting a variety of genomic studies. However, patients who volunteer their genomic data face the risk of privacy invasion. The computation procedures and results either through data outsourcing or data sharing on BDS should not contain any sensitive information of individuals.

\textbf{Data Privacy during Searching}\label{data searching} 
 Searching over encrypted data is required in many scenarios, e.g., users
want to query any data on an untrusted BDS platform without revealing sensitive information of the queried data. The queried data can be either public or anonymous, but the service platform
should not identify its specific content. Furthermore, the search results on big data should not contain any sensitive information of individuals.

In Table \ref{Comparison of techniques} we summarize the set of privacy concerns that need to be addressed by the techniques described in Section \ref{Privacy techniques}.

\begin{table*}[tbp]
\centering
\caption{Summurization of existing privacy-preserving techniques for BDS} \label{Comparison of techniques}
\begin{tabular}{ p{.32\linewidth} p{.45\linewidth}   }

\toprule
 
  $\textbf{Data Privacy Requirements}$ & $\textbf{Privacy-preserving Techniques}$  \\ 
\\ \toprule

 Data Privacy During Downloading  & K-anonymity, L-diversity, and T-closeness \\ \toprule
Data Privacy During Computation & $\mathcal{HE}$ ,
$\mathcal{MPC}$ ,
$\mathcal{DP}$, $\mathcal{FPE}$
\\ \toprule
Data Privacy During Searching  & $\mathcal{ORAM}$, $\mathcal{PEKS}$, $\mathcal{SSE}$  for various search, e.g., keyword, range, boolean, and KNN
  \\ 

\bottomrule
\end{tabular}

\end{table*} 

We have three main groups of privacy preservation techniques: cryptographic tools, \emph{Format-Preserving Encryption}, \emph{Differential Privacy} ($\mathcal{DP}$) and \emph{K-anonymity} (with further enhanced version like \emph{L-diversity} and \emph{T-closeness}) that we present in the next section.

\subsection{Data Privacy Techniques}
\label{Privacy techniques}
In this section, we present a comprehensive overview on recent anonymization techniques used for PPDP. Specifically, our review explains anonymization approaches related to the individual privacy protection. 

\heading{\textbf{Format-Preserving Encryption ($\mathcal{FPE}$)}} $\mathcal{FPE}$ \cite{bellare2009format} is designed to encrypt data of some specified format into a ciphertext of identical format, such as the format of equal length as the original data. FPE can be constructed based on symmetric encryption and other formats also be developed, such as date-time \cite{liu2010format} and charactor \cite{li2012format}.

The $\mathcal{FPE}$ scheme is $E_{K}^{N, T}$ on $X_{N}$ where $E$ is the encryption algorithm. $K$ is the encryption key. $\left\{X_{N}\right\}_{N \in N}$ is the collection of domains. $X_{N}$ is a slice that $X=\bigcup_{N} X_{N})$. $N$ is the format space. $T$ is the tweak.   

\heading{\textbf{Differential Privacy ($\mathcal{DP}$)}} \label{DP} Another important privacy-preserving model is \emph{Differential Privacy ($\mathcal{DP}$)} \cite{Dwork06,mcsherry2007mechanism,desfontaines2020sok}. It is considered as equivalent to perturbing the original data and then computing the queries over that modified data. An algorithm is said to be deferentially private if by looking at the output, one cannot tell whether any individual's data was included in the original dataset or not. It consists on introducing a certain amount of random noise
to data queries such that any statistical analysis over
the whole set is significantly close to the real results, but
inference over any data is infeasible.

In $\mathcal{DP}$ analyst are not provided the direct access to the database containing personal information but an intermediary software is allowed between the database and the analyst to protect the privacy. Currently, $\mathcal{DP}$ is becoming a practical privacy-preserving technique and applied in various application scenarios \cite{hassan2019differential,yang2020local,zhao2019survey}.

 Consider we have two databases $D_{1}$ and $D_{2}$ differ in at
most one element. The general of requirement \cite{dwork2008differential} of $\mathcal{DP}$ can be:
$$
\operatorname{Pr}\left[\mathcal{K}\left(D_{1}\right) \in S\right] \leq \exp (\epsilon) \times \operatorname{Pr}\left[\mathcal{K}\left(D_{2}\right) \in S\right]
$$

Where $\mathcal{K}$ is the algorithm applied by the curator when releasing information. The output of it is $transcript$. If it satisfies the requirement, we say that $\mathcal{K}$ gives $\epsilon$-differential privacy if for all data sets $D_{1}$ and $D_{2}$ differing on at most one element, and all $S \subseteq$ Range $(\mathcal{K})$. 

\heading{\textbf{K-anonymity}} \label{K-anonymity} The k-anonymity \cite{sweeney2002k,lefevre2006mondrian,meyerson2004complexity,el2008protecting} property is satisfied in a release of dataset if individuals information contained in this release cannot be distinguished from at least $k - 1$ individuals in this dataset release. 

To achieve K-anonymity, it is required to have at least $k$ individuals in the dataset who share the set of attributes that might become identifying for each individual. However, k-anonymity prevents identity disclosure but not
attribute disclosure. If there is no diversity in the values
of the sensitive attributes, an attacker can easily discover
the sensitive value of an data through a homogeneity attack. In fact, there are two enhancements of k-anonymity:

The definition of K-anonymity is divided into three parts:
\begin{enumerate}
    \item Attributes. Let $\mathrm{B}\left(A_{1}, \ldots, A_{n}\right)$ be a table with a finite number of tuples. The finite set of attributes of $\mathrm{B}$ are $\left\{A_{1}, \ldots, A_{n}\right\}$
    \item Quasi-identifier. Given a population of entities $\mathrm{U},$ an entity-specific table $\mathrm{T}\left(\mathrm{A}_{1}, \ldots, \mathrm{A}_{\mathrm{n}}\right), f_{\mathrm{c}}: \mathrm{U} \rightarrow$
$\mathrm{T}$ and $f_{g}: \mathrm{T} \rightarrow \mathrm{U}^{\prime},$ where $\mathrm{U} \subseteq \mathrm{U}^{\prime}.$ A quasi-identifier of $\mathrm{T},$ written $Q_{T},$ is a set of attributes $\left\{\mathrm{A}_{\mathrm{i}}, \ldots, \mathrm{A}_{\mathrm{j}}\right\} \subseteq\left\{\mathrm{A}_{1}, \ldots, \mathrm{A}_{\mathrm{n}}\right\}$ where: $\exists p_{i} \in \mathrm{U}$ such that $f_{g}\left(f_{c}\left(p_{i}\right)\left[Q_{T}\right]\right)=p_{i}$
    \item K-anonymity. Let $\operatorname{RT}\left(A_{1}, \ldots, A_{n}\right)$ be a table and $Q I_{R T}$ be the quasi-identifier associated with it. RT is said to satisfy $k$-anonymity if and only if each sequence of values in $\mathrm{RT}\left[Q I_{R T}\right]$ appears with at least $k$ occurrences in $\mathrm{RT}\left[Q I_{R T}\right]$
\end{enumerate}

 \heading{\textbf{L-diversity}}  \label{L-diversity} 
 The L-diversity model is an extension of the k-anonymity model which reduces the granularity of data representation using techniques including generalization and suppression such that any given record maps onto at least k-1 other records in the data. An equivalence class is said to have l-diversity if there are at least l “well-represented” values for the sensitive attribute. A table is said to have l-diversity if every equivalence class of the table has l-diversity \cite{Machanavajjhala:2007,xiao2010hardness}. Compared with k-anonymity, l-diversity can prevent Homogeneity Attack and Background Knowledge Attack. Meanwhile, it also enable more applications in the early like \cite{liu2009query}.
     
L-diversity extends the concept of k-anonymity. Besides the three definition from it, L-diversity has the definitions:

\begin{itemize}
    \item Domain Generalization: A domain $D^{\star}=\left\{P_{1}, P_{2}, \ldots\right\}$ is a generalization (partition) of a domain $D$ if $\cup P_{i}=D$ and $P_{i} \cap P_{j}=\emptyset$ whenever $i \neq j .$ For $x \in D$ we let $\phi_{D^{*}}(x)$ denote the element $P \in D^{\star}$ that contains $x$.
    \item Lack of Diversity. Lack of diversity in the sensitive attribute manifests itself as follows:
$$
\forall s^{\prime} \neq s, \quad n_{\left(q^{*}, s^{\prime}\right)} \ll n_{\left(q^{*}, s\right)}
$$
    \item L-Diversity. A $q^{\star}$-block is L-diverse if it contains at least L well-represented values for the sensitive attribute $S$. A table is L-diverse if every $q^{\star}$-block is L-diverse.
\end{itemize}

The principles of well-present are:

\begin{description}

    \item[Distinct $\ell$ -diversity] The simplest understanding of "well represented" would be to ensure there are at least $\ell$ distinct values for the sensitive attribute in each equivalence class. Distinct $\ell$ -diversity does not prevent probabilistic inference attacks. An equivalence class may have one value appear much more frequently than other values, enabling an adversary to conclude that an entity in the equivalence class is very likely to have that value. This motivated the development of the following two stronger notions of $\ell$ -diversity.
    \item Entropy $\ell$ -diversity. The entropy of an equivalence class $E$ is defined to be
$$
\text {Entropy}(E)=-\sum_{s \in S} p(E, s) \log p(E, s)
$$
A table is said to have entropy $\ell$ -diversity if for every equivalence class $E,$ Entropy$(E) \geq \log \ell .$ Entropy $\ell$ diversity is strong than distinct $\ell$ -diversity. As pointed out in $[12],$ in order to have entropy $\ell$ -diversity for each equivalence class, the entropy of the entire table must be at least $\log (\ell) .$ Sometimes this may too restrictive, as the entropy of the entire table may be low if a few values are very common. This leads to the following less conservative notion of $\ell$ -diversity.
 \item[Recursive $(c, \ell)$ -diversity] Recursive $(c, \ell)$ -diversity makes sure that the most frequent value does not appear too frequently, and the less frequent values do not appear too rarely. Let $m$ be the number of values in an equivalence class, and $r_{i}, 1 \leq i \leq m$ be the number of times that the $i^{t h}$ most frequent sensitive value appears in an equivalence class $E .$ Then $E$ is said to have recursive $(c, \ell)$ -diversity if $r_{1}<c\left(r_{l}+r_{l+1}+\ldots+r_{m}\right)$ A table is said to have recursive $(c, \ell)$ -diversity if all of its equivalence classes have recursive $(c, \ell)$ -diversity.
\end{description}

where $q^{\star}$ the quasi-identifier attribute from generalized domain.

$S$ is the domain of the sensitive attribute.

$p(E, s)$ is the fraction of records in $E$ that have sensitive value $s$.

$n_{\left(q^{*}, s^{\prime}\right)}$ is the umber of tuples $t^{\star}$ in the anonymized table $T^{\star}$ such that $t^{\star}[S]=s$ and $t^{\star}\left[Q^{\star}\right]=q^{\star}$.

\heading{\textbf{T-closeness}} \label{T-closeness}
  Since L-diversity is insufficient to prevent attribute disclosure, researchers find two attack Skewness Attack and Similarity Attack and propose T-closeness model \cite{Li07,rebollo2009t}. An equivalence class is said to have T-closeness if the distance between the distribution of a sensitive attribute in this class and the distribution of the attribute in the whole table is no more than a threshold $t$. A table is said to have T-closeness if all equivalence classes have T-closeness. It is regarded as a more advantageous than the first two techniques and has more general applications like data publishing \cite{li2009closeness,soria2013differential}, data anonymization \cite{domingo2015t}, randomization \cite{rebollo2009t}. 
T-closeness use Earth Mover’s distance (EMD) to measure the distance between the distribution of a sensitive attribute:

$$
\mathrm{D}[\mathbf{P}, \mathbf{Q}]=W O R K(\mathbf{P}, \mathbf{Q}, F)=\sum_{i=1}^{m} \sum_{j=1}^{m} d_{i j} f_{i j}
$$

Then we have two conclusion:
\begin{enumerate}
   
    \item If we have $
    0 \leq d_{i j} \leq 1$ for all $i, j$, then $0 \leq \mathrm{D}[\mathrm{P}, \mathrm{Q}] \leq 1
    $
    
    It means that if the ground distances are normalized, i.e., all distances are between 0 and $1,$ then the EMD between any two distributions is between 0 and $1$. This gives a range from which one can choose the $t$ value for $t$-closeness.
    \item Given two equivalence classes $E_{1}$ and $E_{2}$, let $\mathrm{P}_{1}$, $\mathrm{P}_{2},$ and $\mathrm{P}$ be the distribution of a sensitive attribute in $E_{1}$, $E_{2},$ and $E_{1} \cup E_{2},$ respectively. Then 
    $$
\mathrm{D}[\mathbf{P}, \mathbf{Q}] \leq \frac{\left|E_{1}\right|}{\left|E_{1}\right|+\left|E_{2}\right|} \mathrm{D}\left[\mathbf{P}_{1}, \mathbf{Q}\right]+\frac{\left|E_{2}\right|}{\left|E_{1}\right|+\left|E_{2}\right|} \mathrm{D}\left[\mathbf{P}_{2}, \mathbf{Q}\right]
$$

It follows that $D[\mathbf{P}, \mathbf{Q}] \leq \max \left(D\left[\mathbf{P}_{1}, \mathbf{Q}\right], D\left[\mathbf{P}_{2}, \mathbf{Q}\right]\right) .$ This
means that when merging two equivalence classes, the maximum distance of any equivalence class from the overall distribution can never increase. Thus T-closeness is achievable for any $t \geq 0$
\end{enumerate}

The above fact entails that T-closeness with EMD satisfies the following two requirements.
\begin{description}

    \item[Generalization Property]. Let $\mathcal{T}$ be a table, and let $A$ and $B$ be two generalizations on $\mathcal{T}$ such that $A$ is more general than $B$ If $\mathcal{T}$ satisfies T -closeness using $B,$ then $T$ also satisfies T -closeness using $A$.
    \item[Subset Property]. Let $\mathcal{T}$ be a table and let $C$ be a set of attributes in $\mathcal{T}$. If $\mathcal{T}$ satisfies T-closeness with respect to $C$, then $T$ also satisfies T-closeness with respect to any set of attributes $D$ such that $D \subset C$

\end{description}

\subsection{User Privacy Techniques}

In this section, we provide techniques to achieve the user privacy notion, defined in Section \ref{Introduction}. We argue that, to ensure user privacy, the BDS should be enhanced by other cryptographic techniques that we will be
describe in this section.

\heading{\textbf{Group Signature}} \label{Group Signature}  A group signature \cite{Chaum91} is non-interactive construction
for proving that the data sharer (here the signer) of a certain big data belongs to some group without revealing its identity, which provide anonymity for a data user. In some systems these functionalities are separated and given to a membership manager and revocation manager respectively. Notably, revocable \cite{0eedcc0a721243da97c7426df1798b61,Essam15}, traceable \cite{10.1007/978-3-642-05445-7_6}, or distributed traceable \cite{10.1007/978-3-319-16295-9_18}, or fully dynamic model of \cite{cryptoeprint:2016:368}. Efficient constructions were proposed in \cite{eurocrypt-1991-2127,10.1007/3-540-69053-0_32}, however all of them suffer from the drawback that the size of a public group key and the signatures are proportional to the size of a group. A group signature is called \emph{dynamic}, if the public group key remains unchanged when members join or leave the group or modify their key pairs. The first construction of a dynamic group signature scheme was proposed by Camenisch and Stadler \cite{10.1007/BFb0052252}. Group signature was formalized with concurrent join and an efficient construction by Kiayias and Yung \cite{10.1007/11426639_12}. 

A group signature scheme   $GS=(\mathsf{GKg},\mathsf{GSig},\mathsf{GVf}, \mathsf{Open})$
consists of four polynomial-time algorithms:
\begin{description}

\item[$(\mathsf{gpk}, \mathsf{gmsk}, \mathsf{gsk}) \gets \mathsf{GKg}(1^k,1^n)$.]The randomized group key generation algorithm $\mathsf{GKg}$ takes input $1^k$, $1^n$, where $k \in N$ is the
security parameter and $n \in N$ is the group size (ie. the number of members of the group), and
returns a tuple $(\mathsf{gpk}, \mathsf{gmsk}, \mathsf{gsk})$, where $\mathsf{gpk}$ is the group public key, gmsk is the group manager’s
secret key, and gsk is an n-vector of keys with $\mathsf{gsk}[i]$ being a secret signing key for player $i \in [n]$.
\item[$\sigma \gets \mathsf{GSig}(\mathsf{gsk}i,m))$.] The randomized group signing algorithm $\mathsf{GSig}$ takes as input a secret signing key $\mathsf{gsk}[i]$ and a
message m to return a signature $\sigma$  of m under $\mathsf{gsk}[i]$ ($i \in [n]$).
\item[$(1,0) \gets \mathsf{GVf}(\mathsf{gpk},m,\sigma)$.] The deterministic group signature verification algorithm $\mathsf{GVf}$ takes as input the group public
key $\mathsf{gpk}$, a message m, and a candidate signature $\sigma$ for m to return either 1 or 0.
\item[$(i,\perp) \gets \mathsf{Open}(\mathsf{gmsk},m,\sigma)$.] The deterministic opening algorithm Open takes as input the group manager secret key $\mathsf{gmsk}$,
a message m, and a signature $\sigma$ of m to return an identity i or the symbol $\perp$ to indicate failure.

\end{description}

 \heading{\textbf{Ring Signature}} \label{ring signature}A ring signature scheme \cite{Rivest01} is a group signature scheme but
without group manager to setup a group or revoke a signer's identity.
The formation of a group is spontaneous in the way that group
members can be totally unaware of being integrated to that group. The scheme of Dodis et al. \cite{10.1007/978-3-540-24676-3_36} was the first
to achieve sublinear size signatures in the Random Oracle Model (ROM) \cite{Bellare93}.  

Chow et al. and Bender et al. \cite{10.1145/1128817.1128861,10.1007/11681878_4} simultaneously proposed ring signatures in the standard model. Malavolta and Schroder \cite{10.1007/11681878_4} build setup free and
constant size ring signatures assuming hardness of a variant of the knowledge of
exponent assumption.
\labelText{Ring signature}{RS}

A ring signature scheme is a triple $(\keygen, \mathsf{Sig}, \mathsf{Ver})$.

\begin{description}

    \item[$(x, y) \gets \keygen(1^k)$]  is a probabilistic algorithm which takes security parameter k and outputs private key x and public key y. 
   \item[$\sigma \gets \mathsf{Sig}(1^k, 1^n, x, L, m)$]  is a probabilistic algorithm which takes security parameter k, group size n, private key x, a list $\mathcal{L}$ of n public keys which
includes the one corresponding to x and message m, produces a signature $\sigma$.
\item[$1/0 \gets \mathsf{Ver}(1^k, 1^n, L, m, \sigma)$]  is a boolean algorithm which accepts as inputs
security parameter k, group size n, a list L of n public keys, message m and signature $\sigma$, returns 1 or 0 for accept or reject, respectively. We require that
for any message m, any $(x, y) \gets  \mathsf{Gen}(1^k)$ and any L that includes y, $\mathsf{Ver}(1^k, 1^n, L, m,\mathsf{Sig}(1^k, 1^n, x, L, m)) = 1$.

\end{description}

\heading{\textbf{Attribute-based Signature ($\mathcal{ABS}$)}}  In $\mathcal{ABS}$ \cite{Maji11}, data sharer signs data
with any predicate of their attributes issued from an attribute authority.
\labelText{ABS}{ABS} Let $\mathcal{U}$ be universe of possible attributes. $\Gamma$ is a  claim-predicate over $\mathcal{U}$ which is a boolean function. We say that an attribute set $\vec{x} \subseteq \mathcal{U}$ satisfies $\Gamma$ if $\Gamma(\vec{x})=1$. An  $\mathcal{ABS}$ scheme consists of four algorithms : $\mathsf{Setup},\mathsf{KeyGen}, \mathsf{Sign},\mathsf{Verif}$, which is parameterized by a universe
of possible attributes $\mathcal{U}$ and message space $\mathcal{M}$. 
\begin{description}
\item[$\mathsf{msk}^* \gets \mathsf{Setup}(1^\lambda)$] The attribute-issuing authority $\mathcal{A}$ runs this algorithm. It takes as input the security parameter $1^\lambda$ and outputs a master secret key $\mathsf{msk}^* := (\mathsf{msk}, \mathsf{msk}')$. We call $\mathsf{msk}$  the master secret signing component and $\mathsf{msk}'$ the master secret verification component.  

\item[$\mathsf{sk}_{\vec{x}} \gets \mathsf{KeyGen}(\mathsf{msk},\vec{x} \subseteq \mathcal{U})$] $\mathcal{A}$ runs this randomized algorithm. The $\mathsf{KeyGen}$ algorithm takes as input the master secret signing component $\mathsf{msk}$ with a set of attribute $\vec{x}$. It outputs a secret key $\mathsf{sk}_{\vec{x}}$ corresponding to $\vec{x}$. 

\item[$\sigma \gets \mathsf{Sign}(\mpk, \sk_{\vec{x}},m \in \M, \Gamma)$] The signer $\mathcal{S}$ runs this algorithm. It takes as input a message $m$, the secret key $\mathsf{sk}_{\vec{x}}$ where $\Gamma(\vec{x})=1$, outputs $\sigma$.

\item[$(0,1) \gets \mathsf{Verif}(\mpk, m, \Gamma, \sigma)$] Outputs either $\mathsf{Ac}=0$ or $\mathsf{Rej}=1$.

\end{description}

\begin{table*}[tbp]
\centering
\caption{Comparison of existing  security and privacy techniques}
\label{Taxonomy of existing  security and privacy techniques}

\scalebox{0.60}{\begin{tabular}{p{.23\linewidth} |p{.15\linewidth} p{.43\linewidth} p{.44\linewidth}}

\toprule

\textbf{S\&P Requirements}   &  \textbf{S\&P techniques} &  \textbf{Advantages}&  \textbf{Drawbacks}\\
\toprule
  
Confidentiality of Data &  $\mathcal{IBE}$ & \makecell[l]{Complete access over all resources} & \makecell[l]{PKG knows the secret key users\\Data must be downloaded and decrypted} \\ \\

&$\mathcal{ABE}$ & \makecell[l]{More complex access control\\
on decryption operation than $\mathcal{IBE}$} &  \makecell[l]{ High computation cost \\Data must be downloaded and decrypted} \\\\

&$\mathcal{PRE}$ & \makecell[l]{Delegating decryption rights \\Can be deployed in $\mathcal{IBE}$ or$\mathcal{ABE}$ scheme} &  \makecell[l]{Average computational overhead \\ Data must be downloaded and decrypted} \\\\

&$\mathcal{HE}$                        & \makecell[l]{Securely data outsourcing \\ Sensitive data operations}   & \makecell[l]{Inefficient\\ But $\mathcal{SHE}$ and $\mathcal{PHE}$ are usable} \\ \\

&$\mathcal{ABHE}$ & \makecell[l]{Operation on encrypted data\\Confidentiality\\Access control} &  \makecell[l]{Computational overhead is very high} \\ \midrule
 
Integrity & $\mathcal{DS}$ & \makecell[l]{Provides the non-repudiation} & \makecell[l]{Don't prevent the replay attack \\Slower than MAC} \\\\
   
& $\mathcal{MAC}$ & \makecell[l]{Efficient \\  Suitable for lightweight devices} & \makecell[l]{Does not provide the non-repudiation \\ It is not publicly verifiable.\\ Establishment of Shared Secret}  \\\\

& $\mathcal{PDP}$ &\makecell[l]{Easy and reliable test of data integrity} &  High computation and communication cost  \\\\

& $\mathcal{PS}$  &\makecell[l]{Delegating signing rights \\ Provides the non-repudiation}   &  Unlimited signing rights to the proxy signer \\ \\
   
& $\mathcal{AE}$ [\ref{AE}]& \makecell[l]{Many efficient $\mathcal{AE}$ \\  modes have been developed}   &    \makecell[l]{Does not provide the non-repudiation} \\\midrule

Data Privacy during Computation & $\mathcal{PSI}$ & \makecell[l]{Joint analysis on sensitive encrypted data} &  High computation and communication cost \\ 
& $\mathcal{HE}$ & \makecell[l]{Operation on encrypted data \\  it outputs encrypted data} &  High computation and communication cost \\ \\
&$\mathcal{FE}$ & \makecell[l]{Operates on encrypted data} & \makecell[l]{Efficient to evaluate linear and quadratic functions.}  \\ 
& $\mathcal{MPC}$                     &\makecell[l]{ Joint analysis on sensitive raw data  
 \\ Does not require a centralized party }& \makecell[l]{High computation and \\communication cost}  \\ \midrule
Secure Data Outsourcing & 

$\mathcal{ORAM}$& \makecell[l]{Strong level of data privacy
} &  \makecell[l]{Tends to leak information than using FHE} \\

& $\mathcal{VC}$ & \makecell[l]{Provides the integrity \\Publicly verifiable} &   \makecell[l]{Current solutions are not fully practical}
\\\midrule

Data Privacy during Searching
& $\mathcal{SSE}$ & \makecell[l]{More Efficient than using $\mathcal{PEKS}$  } &  \makecell[l]{It is not suitable for
multi-user data sharing scenarios} \\

& $\mathcal{PEKS}$& \makecell[l]{Can be deployed in $\mathcal{IBE}$ and $\mathcal{ABE}$ } &  \makecell[l]{Balancing between query expressiveness and efficiency}
\\\midrule

User Privacy 

 &  Group Signature& \makecell[l]{Provides traceability} &   The requirement of a group manager \\
 & $\mathcal{ABS}$ & \makecell[l]{Access control is based on use's attribute}  &  \makecell[l]{Credential authority to issue attribute\\ certificates} \\\\
 
&  Ring Signature  & \makecell[l]{More flexibility: No group manager, \\ and the dynamics of group choice} &   \makecell[l]{No anonymity-revocation property \\ Hard to manage/coordinate between\\several signers}  \\\\
     
& \makecell[l]{ACS} & Provides anonymity, authentication and accountability& \makecell[l]{Less efficient than $\mathcal{ABS}$ \\ Does not support complex predicates} \\ \midrule

 Data Privacy 

 & \makecell[l]{K-anonymity}              & \makecell[l]{Easy to implement \\ Reidentification is less when the\\ value of k is high} & \makecell[l]{Background knowledge  Attack\\ Homogeneity attacks\\ Long processing time} \\\\
 & L-diversity          & \makecell[l]{Reduce dataset into summary form \\ Sensitive attribute have at most\\same frequency} & \makecell[l]{Similarity Attack\\ Skewness Attack} \\ \\
 & T-closeness               & Prevent data from skewness attack                 & \makecell[l]{Complex computational procedure \\ Utility is damaged when t is very small} \\\\ 
 & $\mathcal{DP}$                   & \makecell[l]{Most suitable for big data \\ Provides strongest privacy\\guarantee} & \makecell[l]{Complex computational procedure data \\ Utility is damaged when t is very small \\ Noise and loss of information} \\ 
\bottomrule

\end{tabular}}
\end{table*}

 \heading{\textbf{Anonymous Credential Systems ($\mathcal{ACS}$)}} \label{ACS}
In $\mathcal{ACS}$ \cite{Camenisch04}, organizations know the users only by pseudonyms. Different pseudonyms of the same user cannot be linked. Yet, an organization issues a credential to a user whom he knows
by a pseudonym. The corresponding user (under certain pseudonym) can prove to any other organisations that
he is the owner of his credential without revealing anything more than the fact that that user owns such a credential.

A basic anonymous credential
system consists of three entities : $\mathsf{Users}$, an $\mathsf{Authority}$, and $\mathsf{Verifiers}$.

An anonymous credential system has tree procedures: $\keygen$, $\emph{Credential Issuing Protocol}$ and $\emph{Credential Proving Protocol}$, as follows:
\begin{description}

\item $\keygen$ is run by Authority, given security parameter $1^k$, outputs a pair of public key and secret-key: $(\textsc{pk}, \textsc{sk})$.
\item Credential Issuing Protocol  is run by user $\mathcal{U}$, on m that U wants to obtain
a certificate for. Examples of m are properties such as "belongs to some University", "is
over the age of 20." or rights such as "can access the secure room". How Auth detects
whether m is valid or not with regard to $\mathcal{U}$ is outside this protocol.
U executes the credential issuing protocol for m with Auth by using U’s input
m and Auth’s $\textsc{sk}$s. At the end of the protocol, $\mathcal{U}$ obtains a credential Cred,
corresponding to m.
\item Credential Proving Protocol  After $\mathcal{U}$ obtains the credential of m, $\mathcal{U}$ executes the
credential proving protocol of m with a verifier $\mathcal{V}$, that proves $\mathcal{U}$'s possession of Cred.
At the end of the protocol, $\mathcal{V}$ outputs accept if $\mathcal{U}$ really has a valid Cred, otherwise
outputs reject.
\end{description}

\section{Comparison of existing security \& privacy techniques}
\label{Taxonomy security and privacy techniques}
This section presents a comparison of existing  security and privacy techniques for BDS that we will classify on the basis of fulfilled security and privacy requirements, their advantages and drawbacks as shown in Table \ref{Taxonomy of existing  security and privacy techniques}.
To achieve security and privacy in a complex BDS system that needs to meet multiple security and
privacy requirements with desired features, we would like to mention the following three notes:
 \circled{1} No single techniques is a a universal remedy for security and privacy of BDS. Therefore, the appropriate security and privacy techniques (or a combination of them) should be chosen with respect to the security and privacy
requirements and the context of BDS application. \circled{2}  There is no technique that has no side effects
or is perfect in all aspects. When we add a new technique to such complex system, it usually raises new types of attacks or problems. \circled{3} There is always a trade-off between security, privacy and efficiency to make.
\section{Challenging issues \& Future Directions  }
\label{Future Directions}

Although security and privacy techniques have been studied for many years, its implementation and practical adoption is still some unaddressed challenging issues.
 In this section, we discuss
some challenge issues and research directions for security and privacy in BDS :

\textbf{Private Key Management.} Recently, due to the increase in the volume and types of data processed in cloud environments, techniques that allow easy access to Big Data stored in heterogeneous devices in different network environments are emerging security issues for Big Data, Jeong and Shin \cite{10.1007/s11277-015-2990-1} have explained different approaches for key management in big data context such as MPC, server-aided approach, and encryption with signature. In key management with Threshold Signature Scheme, data sharers do not need to keep any key on their own, but instead, they have to share secrets among multiple servers. Keys can be reconstructed using a minimum defined number of secrets using SSS.

\textbf{Balancing Data Sharing Value and Privacy.} To protect the indidividual privacy inside the data, the privacy-preserving techniques such as anonymizing data using whether masking or de-identification techniques are used. However, it's such a double-edged sword: on the one hand it protects the sensitive information inside data such as personal health information (PHI) from disclosure, on the other hand data will lose its quality and would not be enough accurate for analysis anymore. Therefore coming up with a balance between the privacy-protection solutions (anonymization, sharing agreement, and security controls) and accurate data for analyse is essential to be able to access a data that is usable for analytics.

\textbf{Improving Efficiency of Existing Solutions.} Recent cryptographic methods such as private set intersection and homomorphic encryption are powerful cryptographic primitives that have been deployed to solve many security and privacy issues. While these schemes have relatively
good communication cost, the running time can be prohibitive when the set sizes become large due to the
need to perform modular exponentiation for every item in both sets several times. Understanding and balancing these trade-offs theoretically and empirically is a considerable challenge for performing secure BDS. 

\begin{figure}[!htbp]
    \centering
    \includegraphics[width=.6\linewidth]{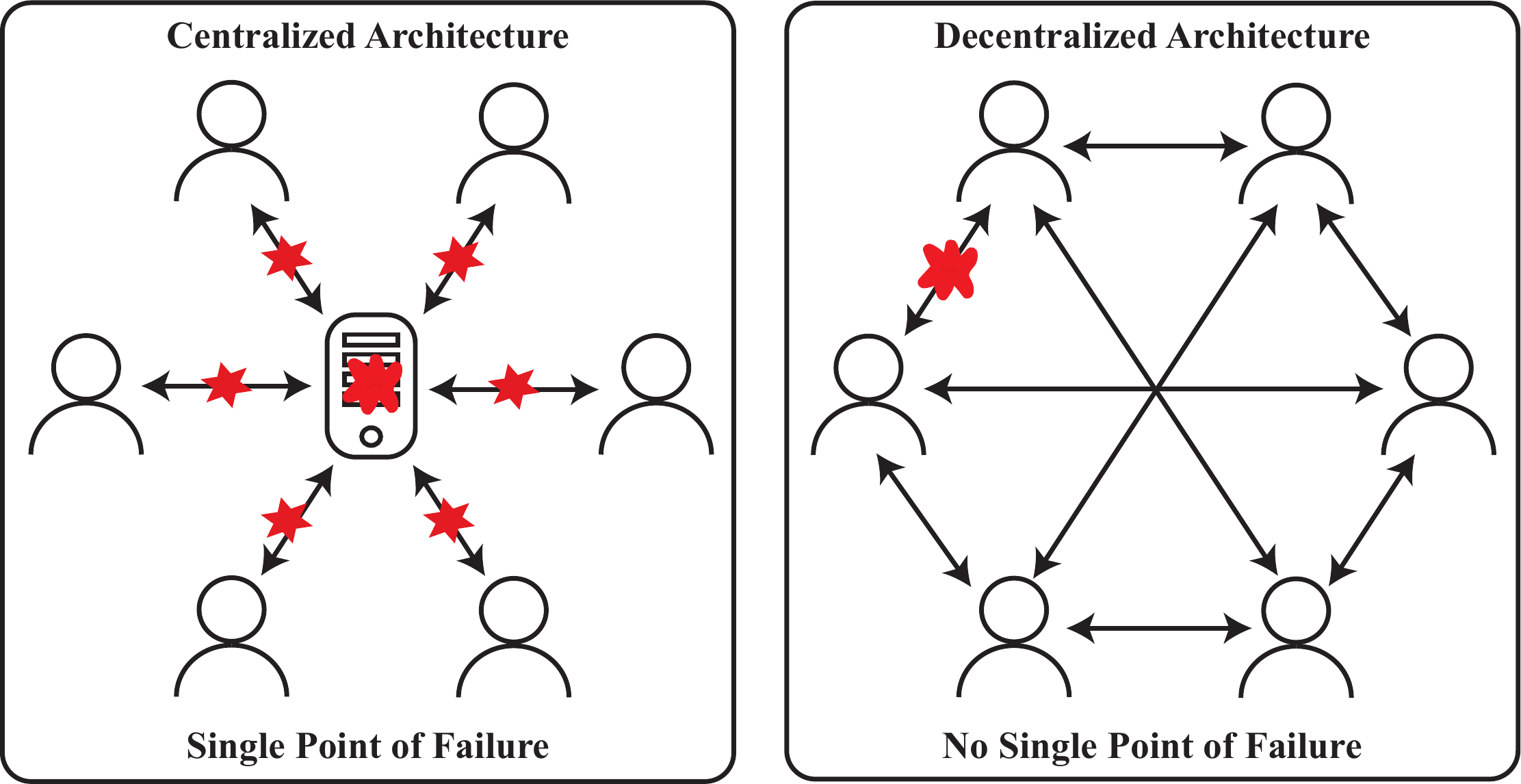}
    \caption{Centralized vs. decentralized networks}
    \label{Centralized vs. decentralized networks}
\end{figure}

\textbf{Eliminating Single Point of Failure: From Centralization to Decentralization, Blockchain as a Solution.} Blockchain technologies is a form of \emph{Distributed Leger Technologies} ($\mathcal{DLT}$s) that provide decentralized platforms which eliminate the need of a single trusted third party (a central authority) and thus get rid of the well-known \emph{Single Point of Failure} ($\mathcal{SPOF}$) issue (which means if the central node goes down, the entire network becomes nonfunctional). This issue provides a breeding ground for cybercriminals as they can target the massive centralized data storage servers vis DDoS, DoS attacks as illustrated in Fig. \ref{Centralized vs. decentralized networks}.

Blockchain is enabled by a combination of technologies such as: peer-to-peer networks, consensus-making, cryptography, smart contract and market mechanisms, among others linking the records (blocks) of the ledger.
The main properties of blockchain are as follows:

\begin{itemize}
     
\item Transparency: The chain is exportable to anywhere and can be downloaded and viewed over the internet.
\item Immutability: Once data is in the chain it cannot be tampered with or altered.

\item Decentralisation: No single entity controls what goes into the chain.

\end{itemize}
  
\textbf{How Blockchain will improve Data Sharing Security?} The combination of blockchain technology and BDS would allow numerous interesting opportunities to improve its security and privacy (as summarized in Figure ~\ref{BC over DB}): 

\begin{figure}[!htbp]
   \centering
   \includegraphics[width=.6\linewidth]{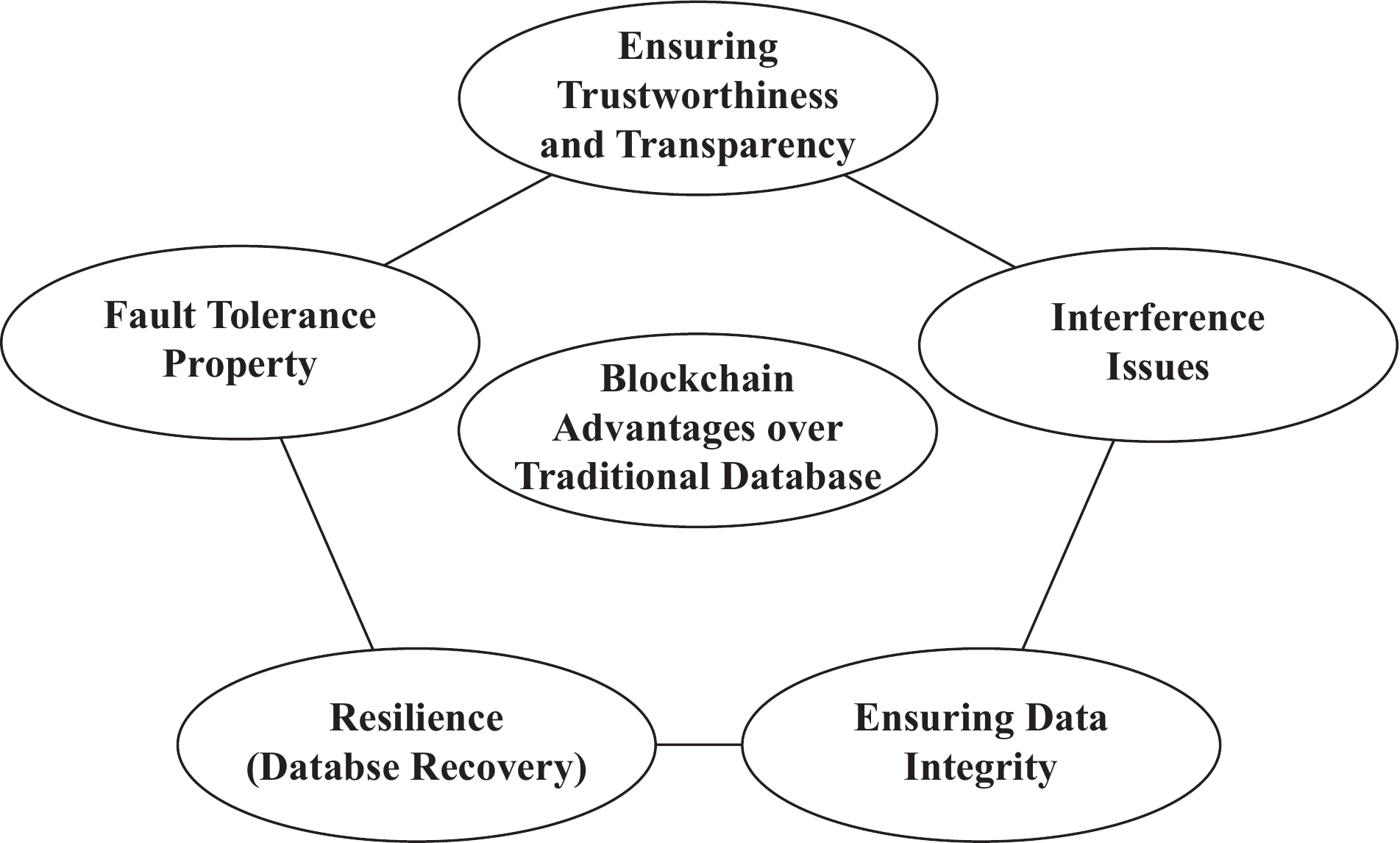}
   \caption{Blockchain advantages over traditional database}
   \label{BC over DB}
\end{figure}

\circled{1} it provides fault tolerance property, which means the distributed nature of blockchain removes the single point of failure; All data is therefore distributed between the nodes of the network. If something is added, edited or deleted in any computer, it will be reflected in all the computers in the network.  Contrary to centralized-based BDS, in case of database failures, the total system of centralized big data is suspended; in blockchain when one workstation goes down, the system will continue operation even with less processing power.\circled{2} it provides database Recovery which means replication of data automatically helps in data recovery if database in any site is damaged.
\circled{3} No interference between users when accessing, sharing and manipulation BDS.\circled{4} Contrary to traditional BDS systems, with blockchain anyone can track data from the source to the end. There are no week points for data to altered or tampered with.
 Participants of network have access to
the holdings and transactions. Using an explorer equipped
with a user's public address is enough to perform their
transactions and actions. \circled{5} The majority of participants in a blockchain system need to approve the transactions and agree upon which data is added, modified or delated. 

Blockchain ensures trust of data by maintaining a decentralized ledger. Data recorded on the blockchain are trustworthy because they must have gone through a verification process which ensures its quality. Data integrity is ensured when details of the origin and interactions concerning a data block are stored on the blockchain and automatically verified (or validated) before it can be acted upon. It also provides for transparency, since activities and transactions that take place on the blockchain network can be traced.

With recent research progress, the
future development direction of  Blockchain Technology includes the following aspects:

\begin{enumerate}
    \item  \heading{DDoS attacks} A denial-of-service attack refers to as the DoS attack on a host. It is the type of cyber-attacks that
disrupt the hosted Internet services by making the host machine or the network resource on the
host unavailable to its intended users. DoS attacks attempt to overload the host system or the host
network resource by flooding with superfluous requests, consequently stalling the fulfillment of
legitimate services.

 \item \heading{Linkability} Different from  user privacy, users should require that the transactions related to themselves cannot be linked.  Because user behaviors in blockchain are traceable, blockchain systems need to protect the transaction privacy of users. 

 \item \heading{The majority 51 \% consensus attack} If an attacker were to take control of 51\% or more of the nodes comprising the blockchain network, the consensus mechanism could be overridden allowing for double spending.

 \item \heading{Private key theft} Transactions in any Blockchain system are authenticated by digital signatures. For example in cryptocurrency context, if Alice wants to send Bob some money, she should sign a transaction by her private key to say “Pay this coin, C to Bob”. However, once a Alice’s private key is lost or stolen, it cannot be recovered. Consequently, the user’s blockchain account can be tampered by others.  

 \item \heading{Irreversibility} Blockchain cannot go back as data is immutable, that cannot always be regarded as a positive aspect, but can quickly turn out to be a major problem in the event of accidents, faulty transactions or fraudulent exchange of goods. Also, such feature does not meet the  "Right to be Forgotten" which gives individuals the right to request that their personal data be removed from a record that is regarded as an important concept in data privacy.

 \item \heading{Selfish-mining attack} One popular use case related to fairness in block mining in
Proof-of-Work (PoW) blockchains, which intuitively requires that a node's mining rewards be proportional to its relative computational power. That is, no node should be able to mine selfishly 
to obtain more rewards than its fair share.
\end{enumerate}

 \textbf{How to secure against unauthorized data re-sharing attack?}  BDS unauthorized data resharing preservation is well known to be impossible to achieve. Despite this limitation, many approaches provide solutions of practical interest by weakening somewhat that requirement. Such approaches include watermarking and copyrighting, that are used as a solution to identify the sharers in re-sharing activity \cite{1381778}. Unfortunately, such approaches cannot prevent unauthorized data re-sharing but only tracking it.

 Using signcryption as a solution to identify the sharers in re-sharing activity is threefold: \circled{1} It provides non-repudiation which provides the victim with transferable evidence against the cheating sharee. \circled{2}) In our construction verifiers (e.g., Smart Contract) will be given the means to determine when the unauthorized user C tries to reshare the A’s data. When this happens, verifiers will be able to contact the data sharer who will provide the proof of his data ownership (this is achieved with the non-repudiation property). \circled{3} It provides authenticated data between different receipts: in order to update the ciphertext recipient, the data owner needs to download it then decrypt the requested data, and further re-signcrypt it under the target user’s public key. This solution is very demanding in terms of computation and communication costs to the data owner which contradicts the motivation of cloud computing. For this reason, to handle this problem in data sharing context, we propose the solution of proxy re-signcryption between different receipts, without compromising the secret key of the data owner. 
 
 Furthermore, the server might ensure that the posted data is not a plaintext but a well encrypted ciphertext in such a way the server can only store the ``encrypted content''. To handle this problem, we introduce a new security notion that we call \emph{verifiable plaintext-aware} that we add to proxy re-signcryption scheme (as described in Fig.  \ref{fig:verifiable plaintext-aware proxy re- signcryption}), which consists of the following steps:

\begin{figure}[!htbp]
\includegraphics[width=.9\linewidth]{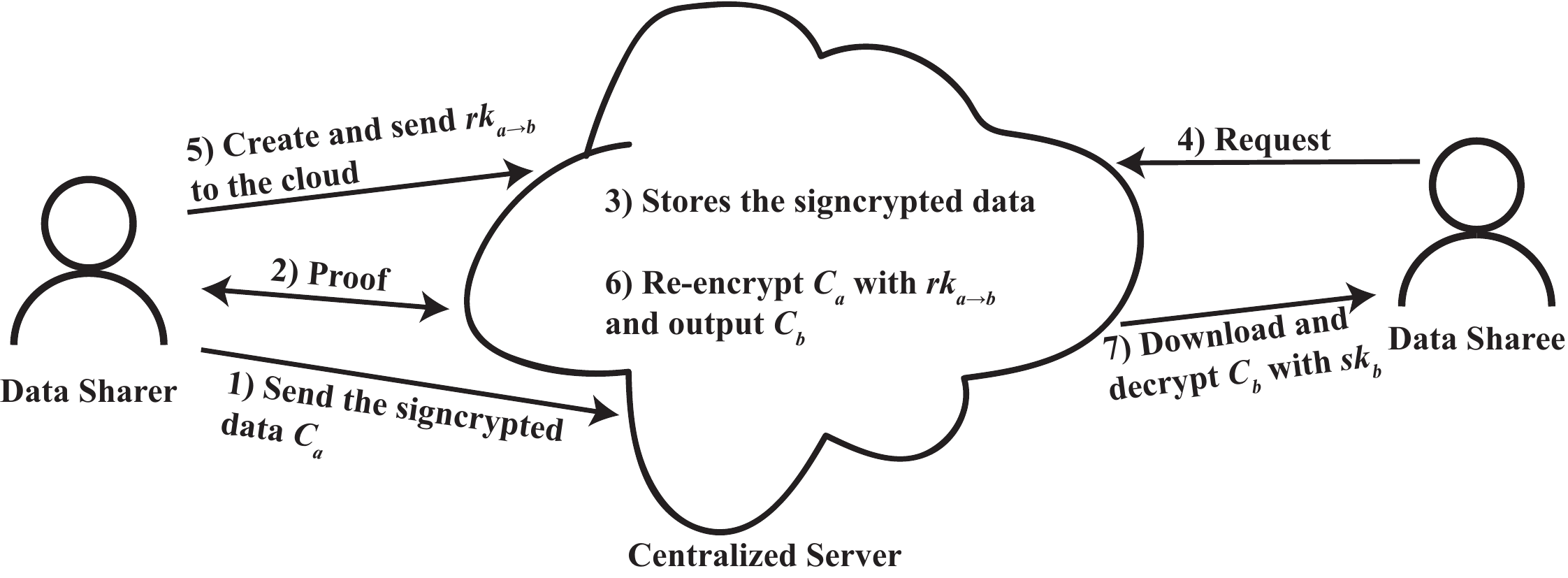}
\caption{System model for verifiable plaintext-aware proxy re-signcryption}
    \label{fig:verifiable plaintext-aware proxy re- signcryption}
\end{figure}

\begin{enumerate}
 \item The data sharer uploads the signcrypted data onto the server.

\item The data sharer convinces the server that the signcrypted data is not a plaintext but a well signcrypted file.

\item If so, the server stores the signcrypted file $C_a$.
\item The data sharee requests decryption delegation with his public key $pk_b$.
\item The data sharee creates and sends the transformation key $rk_{a \rightarrow b}$ to the server.
\item The server re-signcrypt $C_a$ with $rk_{a \rightarrow b}$ and outputs $C_b$.
\item The data sharee download and decrypt $C_b$ with $sk_b$ to recover the original data.

Now let consider three entities: Data Sharer A, Data Sharee B and Unauthorized User C.
We introduce a novel form of unauthorized data resharing prevention called legal unauthorized re-sharing preserving defined as follows:

\end{enumerate}
\section{Conclusion}
\label{conclusion}
 
We have presented a survey on security and privacy of BDS with a number of contributions.
First, we characterized the security and privacy requirements of BDS into two broad categories:
fundamental requirements and additional requirements in the context of BDS. Second, we described the security and privacy techniques for achieving these security and privacy requirements in
BDS.  With growing interest of BDS in both academic research and industry, the security and privacy of BDS have attracted huge interests,
it is impossible to design next generation applications without
publishing and executing data driven algorithms. We conjecture that developing light-weight cryptographic algorithms as well as other
practical security and privacy methods will be a key enabling technology in the future development of BDS and its applications.

\bibliographystyle{acm}
\bibliography{refs.bib}

\begin{thebibliography}{100}

\bibitem{CMAC}
Recommendation for block cipher modes of operation: Galois/counter mode (gcm)
  and gmac.
\newblock \url{https://csrc.nist.gov/publications/sp}.

\bibitem{GCM}
Recommendation for block cipher modes of operation: Galois/counter mode (gcm)
  and gmac.
\newblock \url{https://csrc.nist.gov/publications/detail/sp/800-38d/final}.

\bibitem{10.1007/3-540-46035-7_6}
{\sc An, J.~H., Dodis, Y., and Rabin, T.}
\newblock On the security of joint signature and encryption.
\newblock In {\em Advances in Cryptology --- EUROCRYPT 2002\/} (Berlin,
  Heidelberg, 2002), L.~R. Knudsen, Ed., Springer Berlin Heidelberg,
  pp.~83--107.

\bibitem{Arora98}
{\sc Arora, S., and Safra, S.}
\newblock Probabilistic checking of proofs: A new characterization of np.
\newblock {\em J. ACM 45}, 1 (Jan. 1998), 70–122.

\bibitem{10.1145/1315245.1315318}
{\sc Ateniese, G., Burns, R., Curtmola, R., Herring, J., Kissner, L., Peterson,
  Z., and Song, D.}
\newblock Provable data possession at untrusted stores.
\newblock In {\em Proceedings of the 14th ACM Conference on Computer and
  Communications Security\/} (New York, NY, USA, 2007), CCS '07, Association
  for Computing Machinery, p.~598–609.

\bibitem{ateniese2008scalable}
{\sc Ateniese, G., Di~Pietro, R., Mancini, L.~V., and Tsudik, G.}
\newblock Scalable and efficient provable data possession.
\newblock In {\em Proceedings of the 4th international conference on Security
  and privacy in communication netowrks\/} (2008), pp.~1--10.

\bibitem{Ateniese06}
{\sc Ateniese, G., Fu, K., Green, M., and Hohenberger, S.}
\newblock Improved proxy re-encryption schemes with applications to secure
  distributed storage.
\newblock {\em ACM Trans. Inf. Syst. Secur. 9}, 1 (Feb. 2006), 1–30.

\bibitem{Nuttapong11}
{\sc Attrapadung, N., Libert, B., and de~Panafieu, E.}
\newblock Expressive key-policy attribute-based encryption with constant-size
  ciphertexts.
\newblock In {\em Public Key Cryptography -- PKC 2011\/} (Berlin, Heidelberg,
  2011), D.~Catalano, N.~Fazio, R.~Gennaro, and A.~Nicolosi, Eds., Springer
  Berlin Heidelberg, pp.~90--108.

\bibitem{Babai91}
{\sc Babai, L., Fortnow, L., Levin, L.~A., and Szegedy, M.}
\newblock Checking computations in polylogarithmic time.
\newblock In {\em Proceedings of the Twenty-Third Annual ACM Symposium on
  Theory of Computing\/} (New York, NY, USA, 1991), STOC '91, Association for
  Computing Machinery, p.~21–32.

\bibitem{bellare2009format}
{\sc Bellare, M., Ristenpart, T., Rogaway, P., and Stegers, T.}
\newblock Format-preserving encryption.
\newblock In {\em International workshop on selected areas in cryptography\/}
  (2009), Springer, pp.~295--312.

\bibitem{10.1145/168588.168596}
{\sc Bellare, M., and Rogaway, P.}
\newblock Random oracles are practical: A paradigm for designing efficient
  protocols.
\newblock In {\em Proceedings of the 1st ACM Conference on Computer and
  Communications Security\/} (New York, NY, USA, 1993), CCS '93, Association
  for Computing Machinery, p.~62–73.

\bibitem{Bellare93}
{\sc Bellare, M., and Rogaway, P.}
\newblock Random oracles are practical: A paradigm for designing efficient
  protocols.
\newblock In {\em Proceedings of the 1st ACM Conference on Computer and
  Communications Security\/} (New York, NY, USA, 1993), CCS '93, Association
  for Computing Machinery, p.~62–73.

\bibitem{Bellare03}
{\sc Bellare, M., Rogaway, P., and Wagner, D.}
\newblock A conventional authenticated-encryption mode, 2003.

\bibitem{10.1145/62212.62213}
{\sc Ben-Or, M., Goldwasser, S., and Wigderson, A.}
\newblock Completeness theorems for non-cryptographic fault-tolerant
  distributed computation.
\newblock In {\em Proceedings of the Twentieth Annual ACM Symposium on Theory
  of Computing\/} (New York, NY, USA, 1988), STOC '88, Association for
  Computing Machinery, p.~1–10.

\bibitem{10.1007/11681878_4}
{\sc Bender, A., Katz, J., and Morselli, R.}
\newblock Ring signatures: Stronger definitions, and constructions without
  random oracles.
\newblock In {\em Theory of Cryptography\/} (Berlin, Heidelberg, 2006),
  S.~Halevi and T.~Rabin, Eds., Springer Berlin Heidelberg, pp.~60--79.

\bibitem{4223236}
{\sc {Bethencourt}, J., {Sahai}, A., and {Waters}, B.}
\newblock Ciphertext-policy attribute-based encryption.
\newblock In {\em 2007 IEEE Symposium on Security and Privacy (SP '07)\/}
  (2007), pp.~321--334.

\bibitem{Blaze98}
{\sc Blaze, M., Bleumer, G., and Strauss, M.}
\newblock Divertible protocols and atomic proxy cryptography.
\newblock In {\em Advances in Cryptology --- EUROCRYPT'98\/} (Berlin,
  Heidelberg, 1998), K.~Nyberg, Ed., Springer Berlin Heidelberg, pp.~127--144.

\bibitem{10.1007/BFb0054122}
{\sc Blaze, M., Bleumer, G., and Strauss, M.}
\newblock Divertible protocols and atomic proxy cryptography.
\newblock In {\em Advances in Cryptology --- EUROCRYPT'98\/} (Berlin,
  Heidelberg, 1998), K.~Nyberg, Ed., Springer Berlin Heidelberg, pp.~127--144.

\bibitem{10.1007/978-3-540-28628-8_27}
{\sc Boneh, D., and Boyen, X.}
\newblock Secure identity based encryption without random oracles.
\newblock In {\em Advances in Cryptology -- CRYPTO 2004\/} (Berlin, Heidelberg,
  2004), M.~Franklin, Ed., Springer Berlin Heidelberg, pp.~443--459.

\bibitem{10.1007/978-3-540-24676-3_30}
{\sc Boneh, D., Di~Crescenzo, G., Ostrovsky, R., and Persiano, G.}
\newblock Public key encryption with keyword search.
\newblock In {\em Advances in Cryptology - EUROCRYPT 2004\/} (Berlin,
  Heidelberg, 2004), C.~Cachin and J.~L. Camenisch, Eds., Springer Berlin
  Heidelberg, pp.~506--522.

\bibitem{Boneh01}
{\sc Boneh, D., and Franklin, M.}
\newblock Identity-based encryption from the weil pairing.
\newblock In {\em Advances in Cryptology --- CRYPTO 2001\/} (Berlin,
  Heidelberg, 2001), J.~Kilian, Ed., Springer Berlin Heidelberg, pp.~213--229.

\bibitem{10.5555/646766.704155}
{\sc Boneh, D., and Franklin, M.~K.}
\newblock Identity-based encryption from the weil pairing.
\newblock In {\em Proceedings of the 21st Annual International Cryptology
  Conference on Advances in Cryptology\/} (Berlin, Heidelberg, 2001), CRYPTO
  '01, Springer-Verlag, p.~213–229.

\bibitem{10.1007/978-3-642-19571-6_16}
{\sc Boneh, D., Sahai, A., and Waters, B.}
\newblock Functional encryption: Definitions and challenges.
\newblock In {\em Theory of Cryptography\/} (Berlin, Heidelberg, 2011),
  Y.~Ishai, Ed., Springer Berlin Heidelberg, pp.~253--273.

\bibitem{cryptoeprint:2016:368}
{\sc Bootle, J., Cerulli, A., Chaidos, P., Ghadafi, E., and Groth, J.}
\newblock Foundations of fully dynamic group signatures.
\newblock Cryptology ePrint Archive, Report 2016/368, 2016.
\newblock \url{https://eprint.iacr.org/2016/368}.

\bibitem{10.1007/3-540-48910-X_28}
{\sc Cachin, C., Micali, S., and Stadler, M.}
\newblock Computationally private information retrieval with polylogarithmic
  communication.
\newblock In {\em Advances in Cryptology --- EUROCRYPT '99\/} (Berlin,
  Heidelberg, 1999), J.~Stern, Ed., Springer Berlin Heidelberg, pp.~402--414.

\bibitem{10.1007/3-540-69053-0_32}
{\sc Camenisch, J.}
\newblock Efficient and generalized group signatures.
\newblock In {\em Advances in Cryptology --- EUROCRYPT '97\/} (Berlin,
  Heidelberg, 1997), W.~Fumy, Ed., Springer Berlin Heidelberg, pp.~465--479.

\bibitem{Camenisch04}
{\sc Camenisch, J., and Lysyanskaya, A.}
\newblock Signature schemes and anonymous credentials from bilinear maps.
\newblock In {\em Advances in Cryptology -- CRYPTO 2004\/} (Berlin, Heidelberg,
  2004), M.~Franklin, Ed., Springer Berlin Heidelberg, pp.~56--72.

\bibitem{10.1007/BFb0052252}
{\sc Camenisch, J., and Stadler, M.}
\newblock Efficient group signature schemes for large groups.
\newblock In {\em Advances in Cryptology --- CRYPTO '97\/} (Berlin, Heidelberg,
  1997), B.~S. Kaliski, Ed., Springer Berlin Heidelberg, pp.~410--424.

\bibitem{Chandra17}
{\sc {Chandra}, S., {Ray}, S., and {Goswami}, R.~T.}
\newblock Big data security: Survey on frameworks and algorithms.
\newblock In {\em 2017 IEEE 7th International Advance Computing Conference
  (IACC)\/} (2017), pp.~48--54.

\bibitem{10.1145/62212.62214}
{\sc Chaum, D., Cr\'{e}peau, C., and Damgard, I.}
\newblock Multiparty unconditionally secure protocols.
\newblock In {\em Proceedings of the Twentieth Annual ACM Symposium on Theory
  of Computing\/} (New York, NY, USA, 1988), STOC '88, Association for
  Computing Machinery, p.~11–19.

\bibitem{Chaum91}
{\sc Chaum, D., and van Heyst, E.}
\newblock Group signatures.
\newblock In {\em Advances in Cryptology --- EUROCRYPT '91\/} (Berlin,
  Heidelberg, 1991), D.~W. Davies, Ed., Springer Berlin Heidelberg,
  pp.~257--265.

\bibitem{eurocrypt-1991-2127}
{\sc Chaum, D., and van Heyst, E.}
\newblock Group signatures.
\newblock In {\em Advances in Cryptology - EUROCRYPT '91, Workshop on the
  Theory and Application of of Cryptographic Techniques, Brighton, UK, April
  8-11, 1991, Proceedings\/} (1991), vol.~547 of {\em Lecture Notes in Computer
  Science}, Springer, pp.~257--265.

\bibitem{Chen17}
{\sc Chen, H., Laine, K., and Rindal, P.}
\newblock Fast private set intersection from homomorphic encryption.
\newblock In {\em Proceedings of the 2017 ACM SIGSAC Conference on Computer and
  Communications Security\/} (New York, NY, USA, 2017), CCS '17, Association
  for Computing Machinery, p.~1243–1255.

\bibitem{Cheng15}
{\sc Cheng, H., Rong, C., Hwang, K., Wang, W., and Li, Y.}
\newblock Secure big data storage and sharing scheme for cloud tenants.
\newblock {\em Communications, China 12\/} (06 2015), 106--115.

\bibitem{10.1145/293347.293350}
{\sc Chor, B., Kushilevitz, E., Goldreich, O., and Sudan, M.}
\newblock Private information retrieval.
\newblock {\em J. ACM 45}, 6 (Nov. 1998), 965–981.

\bibitem{10.1007/978-3-642-05445-7_6}
{\sc Chow, S. S.~M.}
\newblock Real traceable signatures.
\newblock In {\em Selected Areas in Cryptography\/} (Berlin, Heidelberg, 2009),
  M.~J. Jacobson, V.~Rijmen, and R.~Safavi-Naini, Eds., Springer Berlin
  Heidelberg, pp.~92--107.

\bibitem{10.1145/1128817.1128861}
{\sc Chow, S. S.~M., Wei, V.~K., Liu, J.~K., and Yuen, T.~H.}
\newblock Ring signatures without random oracles.
\newblock In {\em Proceedings of the 2006 ACM Symposium on Information,
  Computer and Communications Security\/} (New York, NY, USA, 2006), ASIACCS
  '06, Association for Computing Machinery, p.~297–302.

\bibitem{10.1007/978-3-319-12280-9_1}
{\sc Clear, M., and McGoldrick, C.}
\newblock Bootstrappable identity-based fully homomorphic encryption.
\newblock In {\em Cryptology and Network Security\/} (Cham, 2014),
  D.~Gritzalis, A.~Kiayias, and I.~Askoxylakis, Eds., Springer International
  Publishing, pp.~1--19.

\bibitem{10.1145/357830.357847}
{\sc Cramer, R., and Shoup, V.}
\newblock Signature schemes based on the strong rsa assumption.
\newblock {\em ACM Trans. Inf. Syst. Secur. 3}, 3 (Aug. 2000), 161–185.

\bibitem{Curtmola06}
{\sc Curtmola, R., Garay, J., Kamara, S., and Ostrovsky, R.}
\newblock Searchable symmetric encryption: Improved definitions and efficient
  constructions.
\newblock In {\em Proceedings of the 13th ACM Conference on Computer and
  Communications Security\/} (New York, NY, USA, 2006), CCS '06, Association
  for Computing Machinery, p.~79–88.

\bibitem{10.5555/2590701.2590705}
{\sc Curtmola, R., Garay, J., Kamara, S., and Ostrovsky, R.}
\newblock Searchable symmetric encryption: Improved definitions and efficient
  constructions.
\newblock {\em J. Comput. Secur. 19}, 5 (Sept. 2011), 895–934.

\bibitem{curtmola2008mr}
{\sc Curtmola, R., Khan, O., Burns, R., and Ateniese, G.}
\newblock Mr-pdp: Multiple-replica provable data possession.
\newblock In {\em 2008 the 28th international conference on distributed
  computing systems\/} (2008), IEEE, pp.~411--420.

\bibitem{desfontaines2020sok}
{\sc Desfontaines, D., and Pej{\'o}, B.}
\newblock Sok: Differential privacies.
\newblock {\em Proceedings on Privacy Enhancing Technologies 2020}, 2 (2020),
  288--313.

\bibitem{1055638}
{\sc {Diffie}, W., and {Hellman}, M.}
\newblock New directions in cryptography.
\newblock {\em IEEE Transactions on Information Theory 22}, 6 (1976), 644--654.

\bibitem{10.1007/978-3-540-24676-3_36}
{\sc Dodis, Y., Kiayias, A., Nicolosi, A., and Shoup, V.}
\newblock Anonymous identification in ad hoc groups.
\newblock In {\em Advances in Cryptology - EUROCRYPT 2004\/} (Berlin,
  Heidelberg, 2004), C.~Cachin and J.~L. Camenisch, Eds., Springer Berlin
  Heidelberg, pp.~609--626.

\bibitem{domingo2015t}
{\sc Domingo-Ferrer, J., and Soria-Comas, J.}
\newblock From t-closeness to differential privacy and vice versa in data
  anonymization.
\newblock {\em Knowledge-Based Systems 74\/} (2015), 151--158.

\bibitem{Changyu13}
{\sc Dong, C., Chen, L., and Wen, Z.}
\newblock When private set intersection meets big data: An efficient and
  scalable protocol.
\newblock In {\em Proceedings of the 2013 ACM SIGSAC Conference on Computer \&
  Communications Security\/} (New York, NY, USA, 2013), CCS ’13, Association
  for Computing Machinery, p.~789–800.

\bibitem{dwork2008differential}
{\sc Dwork, C.}
\newblock Differential privacy: A survey of results.
\newblock In {\em International conference on theory and applications of models
  of computation\/} (2008), Springer, pp.~1--19.

\bibitem{Dwork06}
{\sc Dwork, C., McSherry, F., Nissim, K., and Smith, A.}
\newblock Calibrating noise to sensitivity in private data analysis.
\newblock In {\em Theory of Cryptography\/} (Berlin, Heidelberg, 2006),
  S.~Halevi and T.~Rabin, Eds., Springer Berlin Heidelberg, pp.~265--284.

\bibitem{Khaled13}
{\sc El~Emam, K.}
\newblock {\em Guide to the de-identification of personal health information}.
\newblock CRC Press, 2013.

\bibitem{el2008protecting}
{\sc El~Emam, K., and Dankar, F.~K.}
\newblock Protecting privacy using k-anonymity.
\newblock {\em Journal of the American Medical Informatics Association 15}, 5
  (2008), 627--637.

\bibitem{10.1007/3-540-39568-7_2}
{\sc ElGamal, T.}
\newblock A public key cryptosystem and a signature scheme based on discrete
  logarithms.
\newblock In {\em Advances in Cryptology\/} (Berlin, Heidelberg, 1985), G.~R.
  Blakley and D.~Chaum, Eds., Springer Berlin Heidelberg, pp.~10--18.

\bibitem{erway2015dynamic}
{\sc Erway, C.~C., K{\"u}p{\c{c}}{\"u}, A., Papamanthou, C., and Tamassia, R.}
\newblock Dynamic provable data possession.
\newblock {\em ACM Transactions on Information and System Security (TISSEC)
  17}, 4 (2015), 1--29.

\bibitem{Fang17}
{\sc Fang, W., Wen, X.~Z., Zheng, Y., and Zhou, M.}
\newblock A survey of big data security and privacy preserving.
\newblock {\em IETE Technical Review 34}, 5 (2017), 544--560.

\bibitem{10.1007/978-3-540-24676-3_1}
{\sc Freedman, M.~J., Nissim, K., and Pinkas, B.}
\newblock Efficient private matching and set intersection.
\newblock In {\em Advances in Cryptology - EUROCRYPT 2004\/} (Berlin,
  Heidelberg, 2004), C.~Cachin and J.~L. Camenisch, Eds., Springer Berlin
  Heidelberg, pp.~1--19.

\bibitem{10.1145/1749603.1749605}
{\sc Fung, B. C.~M., Wang, K., Chen, R., and Yu, P.~S.}
\newblock Privacy-preserving data publishing: A survey of recent developments.
\newblock {\em ACM Comput. Surv. 42}, 4 (June 2010).

\bibitem{6686139}
{\sc {Garg}, S., {Gentry}, C., {Halevi}, S., {Raykova}, M., {Sahai}, A., and
  {Waters}, B.}
\newblock Candidate indistinguishability obfuscation and functional encryption
  for all circuits.
\newblock In {\em 2013 IEEE 54th Annual Symposium on Foundations of Computer
  Science\/} (2013), pp.~40--49.

\bibitem{10.1007/3-540-48910-X_9}
{\sc Gennaro, R., Halevi, S., and Rabin, T.}
\newblock Secure hash-and-sign signatures without the random oracle.
\newblock In {\em Advances in Cryptology --- EUROCRYPT '99\/} (Berlin,
  Heidelberg, 1999), J.~Stern, Ed., Springer Berlin Heidelberg, pp.~123--139.

\bibitem{Gentry09}
{\sc Gentry, C.}
\newblock {\em A Fully Homomorphic Encryption Scheme}.
\newblock PhD thesis, Stanford University, Stanford, CA, USA, 2009.

\bibitem{10.5555/1834954}
{\sc Gentry, C.}
\newblock {\em A Fully Homomorphic Encryption Scheme}.
\newblock PhD thesis, Stanford university, Stanford, CA, USA, 2009.

\bibitem{10.1007/978-3-642-20465-4_9}
{\sc Gentry, C., and Halevi, S.}
\newblock Implementing gentry's fully-homomorphic encryption scheme.
\newblock In {\em Advances in Cryptology -- EUROCRYPT 2011\/} (Berlin,
  Heidelberg, 2011), K.~G. Paterson, Ed., Springer Berlin Heidelberg,
  pp.~129--148.

\bibitem{10.1007/11523468_65}
{\sc Gentry, C., and Ramzan, Z.}
\newblock Single-database private information retrieval with constant
  communication rate.
\newblock In {\em Automata, Languages and Programming\/} (Berlin, Heidelberg,
  2005), L.~Caires, G.~F. Italiano, L.~Monteiro, C.~Palamidessi, and M.~Yung,
  Eds., Springer Berlin Heidelberg, pp.~803--815.

\bibitem{10.1007/978-3-642-40041-4_5}
{\sc Gentry, C., Sahai, A., and Waters, B.}
\newblock Homomorphic encryption from learning with errors:
  Conceptually-simpler, asymptotically-faster, attribute-based.
\newblock In {\em Advances in Cryptology -- CRYPTO 2013\/} (Berlin, Heidelberg,
  2013), R.~Canetti and J.~A. Garay, Eds., Springer Berlin Heidelberg,
  pp.~75--92.

\bibitem{Essam15}
{\sc Ghadafi, E.}
\newblock Efficient distributed tag-based encryption and its application to
  group signatures with efficient distributed traceability.
\newblock {\em Progress in Cryptology - LATINCRYPT 2014 8895\/} (2015),
  327--347.

\bibitem{10.1007/978-3-319-16295-9_18}
{\sc Ghadafi, E.}
\newblock Efficient distributed tag-based encryption and its application to
  group signatures with efficient distributed traceability.
\newblock In {\em Progress in Cryptology - LATINCRYPT 2014\/} (Cham, 2015),
  D.~F. Aranha and A.~Menezes, Eds., Springer International Publishing,
  pp.~327--347.

\bibitem{cryptoeprint:2003:216}
{\sc Goh, E.-J.}
\newblock Secure indexes.
\newblock Cryptology ePrint Archive, Report 2003/216, 2003.
\newblock \url{https://eprint.iacr.org/2003/216}.

\bibitem{10.1145/28395.28420}
{\sc Goldreich, O., Micali, S., and Wigderson, A.}
\newblock How to play any mental game.
\newblock In {\em Proceedings of the Nineteenth Annual ACM Symposium on Theory
  of Computing\/} (New York, NY, USA, 1987), STOC '87, Association for
  Computing Machinery, p.~218–229.

\bibitem{10.1145/233551.233553}
{\sc Goldreich, O., and Ostrovsky, R.}
\newblock Software protection and simulation on oblivious rams.
\newblock {\em J. ACM 43}, 3 (May 1996), 431–473.

\bibitem{Vipul06}
{\sc Goyal, V., Pandey, O., Sahai, A., and Waters, B.}
\newblock Attribute-based encryption for fine-grained access control of
  encrypted data.
\newblock In {\em Proceedings of the 13th ACM Conference on Computer and
  Communications Security\/} (New York, NY, USA, 2006), CCS '06, Association
  for Computing Machinery, p.~89–98.

\bibitem{10.1145/1180405.1180418}
{\sc Goyal, V., Pandey, O., Sahai, A., and Waters, B.}
\newblock Attribute-based encryption for fine-grained access control of
  encrypted data.
\newblock In {\em Proceedings of the 13th ACM Conference on Computer and
  Communications Security\/} (New York, NY, USA, 2006), CCS '06, Association
  for Computing Machinery, p.~89–98.

\bibitem{Green06}
{\sc Green, M., and Ateniese, G.}
\newblock Identity-based proxy re-encryption.
\newblock Cryptology ePrint Archive, Report 2006/473, 2006.
\newblock \url{https://eprint.iacr.org/2006/473}.

\bibitem{Green07}
{\sc Green, M., and Ateniese, G.}
\newblock Identity-based proxy re-encryption.
\newblock In {\em Applied Cryptography and Network Security\/} (Berlin,
  Heidelberg, 2007), J.~Katz and M.~Yung, Eds., Springer Berlin Heidelberg,
  pp.~288--306.

\bibitem{hassan2019differential}
{\sc Hassan, M.~U., Rehmani, M.~H., and Chen, J.}
\newblock Differential privacy techniques for cyber physical systems: a survey.
\newblock {\em IEEE Communications Surveys \& Tutorials 22}, 1 (2019),
  746--789.

\bibitem{Ivan03proxycryptography}
{\sc Ivan, A., and Dodis, Y.}
\newblock Proxy cryptography revisited.
\newblock In {\em in Proceedings of the Network and Distributed System Security
  Symposium (NDSS\/} (2003).

\bibitem{Jain16}
{\sc Jain, P., Gyanchandani, M., and Khare, N.}
\newblock Big data privacy: a technological perspective and review.
\newblock {\em Journal of Big Data 3\/} (12 2016).

\bibitem{1381778}
{\sc {Jengnan Tzeng}, {Wen-Liang Hwang}, and {I-Liang Chern}}.
\newblock An asymmetric subspace watermarking method for copyright protection.
\newblock {\em IEEE Transactions on Signal Processing 53}, 2 (2005), 784--792.

\bibitem{10.1007/s11277-015-2990-1}
{\sc Jeong, Y.-S., and Shin, S.-S.}
\newblock An efficient authentication scheme to protect user privacy in
  seamless big data services.
\newblock {\em Wirel. Pers. Commun. 86}, 1 (Jan. 2016), 7–19.

\bibitem{Kat10}
{\sc Katz, J.}
\newblock {\em Digital Signatures}.
\newblock Springer, 2010.

\bibitem{cryptoeprint:2019:739}
{\sc Kavun, E.~B., Mihajloska, H., and Yalcin, T.}
\newblock A survey on authenticated encryption -- asic designer's perspective.
\newblock Cryptology ePrint Archive, Report 2019/739, 2019.
\newblock \url{https://eprint.iacr.org/2019/739}.

\bibitem{6899132}
{\sc {Khoury}, J., {Lauer}, G., {Pal}, P., {Thapa}, B., and {Loyall}, J.}
\newblock Efficient private publish-subscribe systems.
\newblock In {\em 2014 IEEE 17th International Symposium on
  Object/Component/Service-Oriented Real-Time Distributed Computing\/} (2014),
  pp.~64--71.

\bibitem{10.1007/11426639_12}
{\sc Kiayias, A., and Yung, M.}
\newblock Group signatures with efficient concurrent join.
\newblock In {\em Advances in Cryptology -- EUROCRYPT 2005\/} (Berlin,
  Heidelberg, 2005), R.~Cramer, Ed., Springer Berlin Heidelberg, pp.~198--214.

\bibitem{10.1007/3-540-45539-6_9}
{\sc Kushilevitz, E., and Ostrovsky, R.}
\newblock One-way trapdoor permutations are sufficient for non-trivial
  single-server private information retrieval.
\newblock In {\em Advances in Cryptology --- EUROCRYPT 2000\/} (Berlin,
  Heidelberg, 2000), B.~Preneel, Ed., Springer Berlin Heidelberg, pp.~104--121.

\bibitem{lefevre2006mondrian}
{\sc LeFevre, K., DeWitt, D.~J., and Ramakrishnan, R.}
\newblock Mondrian multidimensional k-anonymity.
\newblock In {\em 22nd International conference on data engineering
  (ICDE'06)\/} (2006), IEEE, pp.~25--25.

\bibitem{li2012format}
{\sc Li, M., Liu, Z., Li, J., and Jia, C.}
\newblock Format-preserving encryption for character data.
\newblock {\em Journal of Networks 7}, 8 (2012), 1239.

\bibitem{Li07}
{\sc {Li}, N., {Li}, T., and {Venkatasubramanian}, S.}
\newblock t-closeness: Privacy beyond k-anonymity and l-diversity.
\newblock In {\em 2007 IEEE 23rd International Conference on Data
  Engineering\/} (2007), pp.~106--115.

\bibitem{li2009closeness}
{\sc Li, N., Li, T., and Venkatasubramanian, S.}
\newblock Closeness: A new privacy measure for data publishing.
\newblock {\em IEEE Transactions on Knowledge and Data Engineering 22}, 7
  (2009), 943--956.

\bibitem{liu2009query}
{\sc Liu, F., Hua, K.~A., and Cai, Y.}
\newblock Query l-diversity in location-based services.
\newblock In {\em 2009 Tenth International Conference on Mobile Data
  Management: Systems, Services and Middleware\/} (2009), IEEE, pp.~436--442.

\bibitem{liu2010format}
{\sc Liu, Z., Jia, C., Li, J., and Cheng, X.}
\newblock Format-preserving encryption for datetime.
\newblock In {\em 2010 IEEE International Conference on Intelligent Computing
  and Intelligent Systems\/} (2010), vol.~2, IEEE, pp.~201--205.

\bibitem{Machanavajjhala:2007}
{\sc Machanavajjhala, A., Kifer, D., Gehrke, J., and Venkitasubramaniam, M.}
\newblock l-diversity: Privacy beyond k-anonymity.
\newblock {\em ACM Transactions on Knowledge Discovery from Data (TKDD) 1}, 1
  (2007).
\newblock Artikel Nr. 3.

\bibitem{9298747}
{\sc {Majeed}, A., and {Lee}, S.}
\newblock Anonymization techniques for privacy preserving data publishing: A
  comprehensive survey.
\newblock {\em IEEE Access 9\/} (2021), 8512--8545.

\bibitem{Maji11}
{\sc Maji, H.~K., Prabhakaran, M., and Rosulek, M.}
\newblock Attribute-based signatures.
\newblock In {\em Topics in Cryptology -- CT-RSA 2011\/} (Berlin, Heidelberg,
  2011), A.~Kiayias, Ed., Springer Berlin Heidelberg, pp.~376--392.

\bibitem{10.1145/238168.238185}
{\sc Mambo, M., Usuda, K., and Okamoto, E.}
\newblock Proxy signatures for delegating signing operation.
\newblock In {\em Proceedings of the 3rd ACM Conference on Computer and
  Communications Security\/} (New York, NY, USA, 1996), CCS '96, Association
  for Computing Machinery, p.~48–57.

\bibitem{mcsherry2007mechanism}
{\sc McSherry, F., and Talwar, K.}
\newblock Mechanism design via differential privacy.
\newblock In {\em 48th Annual IEEE Symposium on Foundations of Computer Science
  (FOCS'07)\/} (2007), IEEE, pp.~94--103.

\bibitem{6234849}
{\sc {Meadows}, C.}
\newblock A more efficient cryptographic matchmaking protocol for use in the
  absence of a continuously available third party.
\newblock In {\em 1986 IEEE Symposium on Security and Privacy\/} (1986),
  pp.~134--134.

\bibitem{meyerson2004complexity}
{\sc Meyerson, A., and Williams, R.}
\newblock On the complexity of optimal k-anonymity.
\newblock In {\em Proceedings of the twenty-third ACM SIGMOD-SIGACT-SIGART
  symposium on Principles of database systems\/} (2004), pp.~223--228.

\bibitem{0eedcc0a721243da97c7426df1798b61}
{\sc Nisansala, M., Perera, S., and Koshiba, T.}
\newblock Fully secure lattice-based group signatures with verifier-local
  revocation.
\newblock In {\em Proceedings - 31st IEEE International Conference on Advanced
  Information Networking and Applications, AINA 2017\/} (United States, May
  2017), Institute of Electrical and Electronics Engineers Inc., pp.~795--802.
\newblock 31st IEEE International Conference on Advanced Information Networking
  and Applications, AINA 2017 ; Conference date: 27-03-2017 Through 29-03-2017.

\bibitem{10.1145/100216.100289}
{\sc Ostrovsky, R.}
\newblock Efficient computation on oblivious rams.
\newblock In {\em Proceedings of the Twenty-Second Annual ACM Symposium on
  Theory of Computing\/} (New York, NY, USA, 1990), STOC '90, Association for
  Computing Machinery, p.~514–523.

\bibitem{Ostrovsky90}
{\sc Ostrovsky, R.}
\newblock Efficient computation on oblivious rams.
\newblock In {\em Proceedings of the Twenty-Second Annual ACM Symposium on
  Theory of Computing\/} (New York, NY, USA, 1990), STOC '90, Association for
  Computing Machinery, p.~514–523.

\bibitem{10.1007/3-540-48910-X_16}
{\sc Paillier, P.}
\newblock Public-key cryptosystems based on composite degree residuosity
  classes.
\newblock In {\em Advances in Cryptology --- EUROCRYPT '99\/} (Berlin,
  Heidelberg, 1999), J.~Stern, Ed., Springer Berlin Heidelberg, pp.~223--238.

\bibitem{10.1145/3154794}
{\sc Pinkas, B., Schneider, T., and Zohner, M.}
\newblock Scalable private set intersection based on ot extension.
\newblock {\em ACM Trans. Priv. Secur. 21}, 2 (Jan. 2018).

\bibitem{Pirretti}
{\sc Pirretti, M., Traynor, P., McDaniel, P., and Waters, B.}
\newblock Secure attribute-based systems.
\newblock In {\em Proceedings of the 13th ACM Conference on Computer and
  Communications Security\/} (New York, NY, USA, 2006), CCS '06, Association
  for Computing Machinery, p.~99–112.

\bibitem{poldrack2014making}
{\sc Poldrack, R.~A., and Gorgolewski, K.~J.}
\newblock Making big data open: data sharing in neuroimaging.
\newblock {\em Nature neuroscience 17}, 11 (2014), 1510--1517.

\bibitem{rajendran2017study}
{\sc Rajendran, K., Jayabalan, M., and Rana, M.~E.}
\newblock A study on k-anonymity, l-diversity, and t-closeness techniques.
\newblock {\em IJCSNS 17}, 12 (2017), 172.

\bibitem{rebollo2009t}
{\sc Rebollo-Monedero, D., Forne, J., and Domingo-Ferrer, J.}
\newblock From t-closeness-like privacy to postrandomization via information
  theory.
\newblock {\em IEEE Transactions on Knowledge and Data Engineering 22}, 11
  (2009), 1623--1636.

\bibitem{richey2016global}
{\sc Richey, R.~G., Morgan, T.~R., Lindsey-Hall, K., and Adams, F.~G.}
\newblock A global exploration of big data in the supply chain.
\newblock {\em International Journal of Physical Distribution \& Logistics
  Management\/} (2016).

\bibitem{Rivest1978}
{\sc Rivest, R.~L., Adleman, L., and Dertouzos, M.~L.}
\newblock On data banks and privacy homomorphisms.
\newblock {\em Foundations of Secure Computation, Academia Press\/} (1978),
  169--179.

\bibitem{10.1145/359340.359342}
{\sc Rivest, R.~L., Shamir, A., and Adleman, L.}
\newblock A method for obtaining digital signatures and public-key
  cryptosystems.
\newblock {\em Commun. ACM 21}, 2 (Feb. 1978), 120–126.

\bibitem{Rivest01}
{\sc Rivest, R.~L., Shamir, A., and Tauman, Y.}
\newblock How to leak a secret.
\newblock In {\em Advances in Cryptology --- ASIACRYPT 2001\/} (Berlin,
  Heidelberg, 2001), C.~Boyd, Ed., Springer Berlin Heidelberg, pp.~552--565.

\bibitem{Sahai05}
{\sc Sahai, A., and Waters, B.}
\newblock Fuzzy identity-based encryption.
\newblock In {\em Advances in Cryptology -- EUROCRYPT 2005\/} (Berlin,
  Heidelberg, 2005), R.~Cramer, Ed., Springer Berlin Heidelberg, pp.~457--473.

\bibitem{Shamir85}
{\sc Shamir, A.}
\newblock Identity-based cryptosystems and signature schemes.
\newblock In {\em Advances in Cryptology\/} (Berlin, Heidelberg, 1985), G.~R.
  Blakley and D.~Chaum, Eds., Springer Berlin Heidelberg, pp.~47--53.

\bibitem{Siddique18}
{\sc Siddique, M., Mirza, M.~A., Ahmad, M., Chaudhry, J., and Islam, R.}
\newblock A survey of big data security solutions in healthcare.
\newblock In {\em Security and Privacy in Communication Networks\/} (Cham,
  2018), R.~Beyah, B.~Chang, Y.~Li, and S.~Zhu, Eds., Springer International
  Publishing, pp.~391--406.

\bibitem{article}
{\sc Simplicio, M., Oliveira, B., Margi, C., Barreto, P., Carvalho, T., and
  Naslund, M.}
\newblock Survey and comparison of message authentication solutions on wireless
  sensor networks.
\newblock {\em Ad Hoc Networks 11\/} (05 2013), 1221 -- 1236.

\bibitem{Simplcio2009TheMM}
{\sc Simpl{\'i}cio, M.~A., d'Aquino F.~F. S.~Barbuda, P., Barreto, P. S. L.~M.,
  Carvalho, T., and Margi, C.~B.}
\newblock The marvin message authentication code and the lettersoup
  authenticated encryption scheme.
\newblock {\em Secur. Commun. Networks 2\/} (2009), 165--180.

\bibitem{soria2013differential}
{\sc Soria-Comas, J., and Domingo-Ferrert, J.}
\newblock Differential privacy via t-closeness in data publishing.
\newblock In {\em 2013 Eleventh Annual Conference on Privacy, Security and
  Trust\/} (2013), IEEE, pp.~27--35.

\bibitem{sweeney2002k}
{\sc Sweeney, L.}
\newblock k-anonymity: A model for protecting privacy.
\newblock {\em International Journal of Uncertainty, Fuzziness and
  Knowledge-Based Systems 10}, 05 (2002), 557--570.

\bibitem{Terzi15}
{\sc {Terzi}, D.~S., {Terzi}, R., and {Sagiroglu}, S.}
\newblock A survey on security and privacy issues in big data.
\newblock In {\em 2015 10th International Conference for Internet Technology
  and Secured Transactions (ICITST)\/} (2015), pp.~202--207.

\bibitem{Thangaraj17}
{\sc Thangaraj, M., and Balamurugan, S.}
\newblock Survey on big data security framework.
\newblock In {\em Knowledge Management in Organizations\/} (Cham, 2017),
  L.~Uden, W.~Lu, and I.-H. Ting, Eds., Springer International Publishing,
  pp.~470--481.

\bibitem{Walfish15}
{\sc Walfish, M., and Blumberg, A.~J.}
\newblock Verifying computations without reexecuting them.
\newblock {\em Commun. ACM 58}, 2 (Jan. 2015), 74–84.

\bibitem{Waters11}
{\sc Waters, B.}
\newblock Ciphertext-policy attribute-based encryption: An expressive,
  efficient, and provably secure realization.
\newblock In {\em Public Key Cryptography -- PKC 2011\/} (Berlin, Heidelberg,
  2011), D.~Catalano, N.~Fazio, R.~Gennaro, and A.~Nicolosi, Eds., Springer
  Berlin Heidelberg, pp.~53--70.

\bibitem{welch2016determinants}
{\sc Welch, E.~W., Feeney, M.~K., and Park, C.~H.}
\newblock Determinants of data sharing in us city governments.
\newblock {\em Government Information Quarterly 33}, 3 (2016), 393--403.

\bibitem{wu2019data}
{\sc Wu, H., Cao, J., Yang, Y., Tung, C.~L., Jiang, S., Tang, B., Liu, Y.,
  Wang, X., and Deng, Y.}
\newblock Data management in supply chain using blockchain: Challenges and a
  case study.
\newblock In {\em 2019 28th International Conference on Computer Communication
  and Networks (ICCCN)\/} (2019), IEEE, pp.~1--8.

\bibitem{xiao2010hardness}
{\sc Xiao, X., Yi, K., and Tao, Y.}
\newblock The hardness and approximation algorithms for l-diversity.
\newblock In {\em Proceedings of the 13th International Conference on Extending
  Database Technology\/} (2010), pp.~135--146.

\bibitem{yang2020local}
{\sc Yang, M., Lyu, L., Zhao, J., Zhu, T., and Lam, K.-Y.}
\newblock Local differential privacy and its applications: A comprehensive
  survey.
\newblock {\em arXiv preprint arXiv:2008.03686\/} (2020).

\bibitem{Yao1982}
{\sc Yao, A.~C.}
\newblock Protocols for secure computations.
\newblock In {\em Proceedings of the 23rd Annual IEEE Symposium on Foundations
  of Computer Science\/} (Washington, DC, USA, 1982), IEEE Computer Society,
  pp.~160--164.

\bibitem{zhao2019survey}
{\sc Zhao, P., Zhang, G., Wan, S., Liu, G., and Umer, T.}
\newblock A survey of local differential privacy for securing internet of
  vehicles.
\newblock {\em The Journal of Supercomputing\/} (2019), 1--22.

\bibitem{Zheng97}
{\sc Zheng, Y.}
\newblock Digital signcryption or how to achieve cost(signature \& encryption)
  $\ll$ cost(signature) + cost(encryption).
\newblock In {\em Advances in Cryptology --- CRYPTO '97\/} (Berlin, Heidelberg,
  1997), B.~S. Kaliski, Ed., Springer Berlin Heidelberg, pp.~165--179.

\end{thebibliography}

\end{document}